\documentclass[fleqn,usenatbib]{rasti}

\usepackage{newtxtext,newtxmath}

\usepackage[T1]{fontenc}

\DeclareRobustCommand{\VAN}[3]{#2}
\let\VANthebibliography\thebibliography
\def\thebibliography{\DeclareRobustCommand{\VAN}[3]{##3}\VANthebibliography}

\usepackage{graphicx}	
\usepackage{amsmath}	
\usepackage{multicol}
\usepackage{cuted}
\usepackage{flushend}
\usepackage{caption}
\usepackage{subcaption}
\usepackage{wrapfig,lipsum,booktabs}
\usepackage[flushleft]{threeparttable}

\title[Lunar Far-Side Radio Arrays: A Preliminary Site Survey]{Lunar Far-Side Radio Arrays: A Preliminary Site Survey}

\author[Z. A. Le Conte et al.]{
Zoe A. Le Conte,$^{1,2,3}$\thanks{E-mail: zoe.a.le-conte@durham.ac.uk}
Martin Elvis,$^{1}$
Philipp A. Gläser$^{4}$
\\
$^{1}$Center for Astrophysics | Harvard \& Smithsonian, 60 Garden Street, Cambridge, MA 02138, USA\\
$^{2}$Centre for Extragalactic Astronomy, Department of Physics, Durham University, South Road, Durham DH1 3LE, UK\\
$^{3}$Faculty of Engineering and Physical Sciences, University of Southampton, Southampton, SO17 1BJ, UK\\
$^{4}$Technical University of Berlin, Kaiserin-Augusta-Allee 104-106, Berlin, 10553, Germany
}

\date{Accepted 16/06/2023. Received 25/05/2023; in original form 15/12/2022}

\pubyear{2023}

\begin{document}
\label{firstpage}
\pagerange{\pageref{firstpage}--\pageref{lastpage}}
\maketitle

\begin{abstract}
The origin and evolution of structure in the Universe could be studied in the Dark Ages. The highly redshifted HI signal between $30 < z < 80$ is the only observable signal from this era. Human radio interference and ionospheric effects limit Earth-based radio astronomy to frequencies $> 30$ MHz. To observe the low-frequency window with research from compact steep spectrum sources, pulsars, and solar activity, a 200 km baseline lunar far-side radio interferometer has been much discussed. This paper conducts a preliminary site survey of potential far-side craters, which are few in number on the mountainous lunar far-side. Based on LRO LOLA data, 200 m resolution topographic maps of eight far-side sites were produced, and slope and roughness maps were derived from them. A figure of merit was created to determine the optimum site. Three sites are identified as promising. There is a need to protect these sites for astronomy.
\end{abstract}


\section{Introduction}
\label{sec:intro}
The far-side of the Moon is sheltered from anthropogenic low-frequency radio emissions and is considered the best location for a low-frequency radio telescope to investigate a broad scope of research areas, including the cosmic Dark Ages, compact steep spectrum (CSS) sources and solar radio bursts. This paper seeks to establish the best sites on the far-side for an array that maximally extracts information from the highly redshifted 21 cm signal and surveys the low-frequency radio sky. 

The Astro2020 Decadal Survey of Astronomy and Astrophysics identified the Dark Ages as the discovery era for cosmology \citep[See][]{NAP26141}. Studying the Universe during the so-called Dark Ages can deepen our understanding of the evolution of large-scale structure and the theorised Epoch of inflation (\citet{Adams1993}). The Dark Ages can be probed using the signal from a highly redshifted 21 cm transition of neutral hydrogen. This forbidden transition is produced by the emission of a photon with rest frequency 1420 MHz, by the process of the hydrogen’s proton and electron spin-spin coupling, and results in a hyperfine transition spin-flip and subsequently the emission of a photon (\citet{Muller1951}). 

The 21 cm signal can be measured as a sky-averaged global signal or spatial fluctuations. The 21 cm absorption trough traces the density field of the universe during the Dark Ages epoch. Only one detection of the 21 cm absorption profile, in the Cosmic Dawn, in the sky-averaged spectrum, has been claimed (but this has been disputed (\citet{singh2022})). This detection comes from the \textit{Experiment to Detect the Global Epoch of Reionization Signature} (EDGES) experiment and is centred at 78 ± 1 MHz corresponding to a redshift $z \approx 17$ (\citet{Bowman2018}). The EDGES trough is $\sim3$ times deeper than the standard model, and adiabatic cooling is insufficient to explain the result \citep[see, e.g.,][]{Burns_21, Bowman2018, Mebane2020}.

Spatial fluctuations in the highly redshifted 21 cm signal test the structure formation predictions of the standard cosmological model. Measuring the absorption of galaxy precursor neutral hydrogen gas clouds against the CMB increases the number of observable modes enormously. The billions of galaxies in galaxy surveys were potentially constituted from hundreds of millions of cold hydrogen gas clouds; therefore, trillions of independent modes can be measured (\citet{Silk2018}). They provide unprecedented constraints of the early Universe. In particular, on the matter power spectrum spectral index (\citet{Mao2008}), non-gaussianity (\citet{Munoz2015}) of the inflationary era, the mass of the neutrino and the curvature of the Universe (\citet{Mao2008}). 21 cm observations have two other major advantages over the CMB (\citet{Furlanetto2006}): (i) They are unaffected by Silk damping (the smoothing phenomenon of  primordial density fluctuations) and so spatial fluctuations hold at much smaller mass scales in comparison to the CMB (\citet{Silk1967}); (ii) They can be used to construct the hydrogen structure in a 3D volume (\citet{Furlanetto2006}).

An ultra-long wavelength radio interferometer is required to measure high redshift spatial fluctuations. The 21 cm signal is observed by measuring the neutral hydrogen spin temperature against the CMB radiation temperature. The evolution of the transition spin temperature is explained by \citet{Loeb2004}, limiting the redshift range to $30 < z < 80$. For this redshift range, the 21 cm signal lies in the meters wavelength range (6.5 - 17.0 m) and the tens of MHz frequency range (17.6 - 46.1 MHz). \cite{2021Koopmans} specifies the requirements for a lunar-based instrument to extract the faint 21 cm signal with S/N > 10 \citep[Refer to Section 7 of][and Figure 8 for S/N predictions]{2021Koopmans}{}{}. The cosmological radio interferometer relies on a densely packed (order of $10^{4}$ dipoles) core to reach the sensitivity of the 21 cm signal and a baseline of a few kilometres.

Very low-frequency radio astronomy has been recognized to have several other potential uses described below. Currently, the largest radio telescope, LOFAR (Low-Frequency Array, \cite{Haarlem2013}), operates at the longest wavelengths observed from Earth. The LOFAR Low Band Antenna (LBA) Sky Survey (LoLSS, \cite{Gasperin2021}) observes the frequency range 42 - 66 MHz with high-sensitivity $\sim$ 1 mJy beam$^{-1}$, and high-resolution $\sim$ 15 arcsecs. To reach these exceptional parameters, the baseline of the LBA is $\sim$ 100 km. Baselines up to $\sim$ 200 km can be attained on the lunar far-side, and applying the same imaging capabilities as LOFAR would achieve 10 MHz with an angular resolution of $\sim$ 25 arcsecs. The multi-research capabilities of a lunar far-side radio interferometer would extend the capabilities of LoLSS.

There is a variety of low-frequency radio-loud active galactic nuclei (AGN) sources, from CSS at a few hundred years to radio galaxies on the scale of Mpcs and aged at $\sim$ 1 Gyr \citep[e.g.,][]{Dabhade2020}{}{}. Jets, driven by synchrotron processes, propagate beyond the host galaxy and are classed as Fanaroff and Riley radio galaxies (FRIs and FRIIs, \cite{Fanaroff1974}) with signature double-lobed structures that can span up to hundreds of kiloparsecs. CSS, by contrast, have radio propagation extents $\sim$< 20 kpc and some are thought to grow in strength and evolve into FRIs and FRIIs \citep[e.g.,][]{Fanti1990}{}{}. Alternatively, the phenomenon "frustration" describes the inability of jets in CSS sources to penetrate the interstellar medium (ISM) due to a dense ISM or intrinsically weak jets \citep[e.g.,][]{Van1984}{}{}. High-resolution and sensitive low-frequency measurements of compact objects could assess the extent of the physical processes responsible for the termination of jets \citep[e.g.,][]{McKean2016}.

An ultra-long wavelength interferometer would open a window for new low-frequency pulsars studies \citet{Stappers2011}. Despite pulsars being intrinsically brightest at low radio frequency ranges (< 300 MHz), past surveys were conducted at $\sim$ 350 MHz or $\sim$ 1.4 GHz due to observational biases. \cite{Pilia2016} investigates the evolution of pulsar profiles over a frequency range < 200 MHz with respect to magnetospheric origins and dark matter-induced variations. A specific trend cannot be identified with decreasing radio frequency, and the distribution of the average spectral index is compatible with different pulsar behaviours.

Solar activity, such as coronal mass ejections (CME), solar flares and sunspots, can be observed across the low-frequency radio spectrum; from Type III bursts extending to tens of KHz to the observation of Type I bursts up to hundreds of MHz. Thermal or non-thermal, coherent or incoherent emission mechanisms produce solar radio bursts which trace energetic phenomena in the interplanetary medium and solar corona. Type II radio bursts are emitted from highly energetic accelerated electrons from shock waves driven by supersonic CMEs \citep[e.g.,][]{Kouloumvakos_2021}{}{}. Monitoring solar activity below 10 MHz would advance studies in CME release mechanisms and assess the impacts of CMEs on Earth.

For a radio interferometer applicable to a wide range of research, this paper adopts 200 km as the minimum interferometer baseline distance.

Earth-based low-frequency/long-wavelength radio astronomy has been constrained by the uncontrollable production of human-created radio frequency interference (RFI) over recent decades. Shortwave broadcasting, emitted by telecommunication and radio communication satellites in Earth orbits and ground-based transmitters, floods radio telescope instruments with noise below 30 MHz (e.g., \citet{Maccone2021}). Radio observations are further restricted by ionospheric effects, with frequencies below 10 MHz being absorbed and distorted, becoming effectively opaque (\citet{Kaiser2000}).

A solution is constructing radio astronomy instruments on the far-side of the Moon. The 3,400 km thickness and tidal locking of the Moon shield the far-side surface from RFI by as much as 90 dB and creates a Radio Quiet Zone (RQZ) (e.g., \citet{Kim2021}). The location of this zone is determined by the attenuation of radio waves around the lunar limb. There are three factors to consider: (1) geometry; (2) libration; (3) diffraction. (1) Geometry refers to the RFI produced from the surface of the Earth or by a Geosynchronous Equatorial Orbit (GEO) satellite. Lunar longitude is measured from the sub-Earth point, so the limb as seen from Earth is at $\pm 90$°. The longitude at which RFI from the satellite diminishes to 90 dB is $\sim 6$° beyond the lunar limb. (2) Diffraction further constrains the lunar RQZ. At 10 MHz with a -80 dB threshold, waves are attenuated $\sim 4$° beyond the geometric limit, reducing the RQZ to $\sim 160$° on the lunar far-side \citep[see][]{Bassett2020}. (3) Optical librations occur from a shift in viewing angle due to the non-circular and inclined orbit of the Moon and have three forms: latitude, parallactic, and longitude. Longitude is the most significant libration in which the RQZ varies by $\approx$ 8° (\citet{Ratkowski2014}). Combining these three effects, the -80 dB RQZ in longitude begins at $\pm 108$° stretching 4,366 km across the anti-Earth in longitude and latitude begins at $\pm 79$°, centred at the anti-Earth, spanning 4791 km.

Concepts for radio interferometer arrays on the Moon have been developed. The FARSIDE (Farside Array for Radio Science Investigations of the Dark Ages and Exoplanets) mission concept is a recent example \citep[see][]{Burns2021}. The development of radio interferometer arrays on the Moon will be an evolutionary process, starting with the few-kilometre scale, which has a large choice of locations. Instead, the 200 km-class arrays that are ultimately needed are much harder to find a site. Any such site must be traversable by a rover. Rovers for the Moon and Mars have limited capabilities to deal with steep or rough terrain (Table \ref{tab:rovers}). For example, NASA’s Perseverance is a 26 cm diameter wheeled Mars rover with a 30° safety incline limit. Similarly, NASA's Volatiles Investigating Polar Exploration Rover (VIPER), which will explore the South Pole, has $\sim$> 40\% slip on slopes > 15° and can traverse objects only up to 10 cm in height. These considerations greatly reduce the suitable site selection as the lunar far-side is extremely mountainous, without the large, smooth maria of the near side. 

This paper examines the topography of eight promising sites for a multi-purpose 200 km-scale low-frequency radio telescope on the Moon. The contents of this paper are observations and data reduction in Section \ref{sec:LOLA}; Candidate selection in Section \ref{sec:candidate}; Data analysis methods in Section \ref{sec:Method}; Map products in Section \ref{sec:Results}; Site comparisons in Section \ref{sec:fom}; Discussion in Section \ref{sec:Discussion}; Conclusions in Section \ref{sec:Conclusion}.

\section{Observations and Data Reduction}
\label{sec:LOLA}
\subsection{Lunar Elevation: Lunar Reconnaissance Orbiter}
The data used in this paper was taken by the NASA Lunar Reconnaissance Orbiter (LRO). LRO orbited the Moon in a nominal circular 50 km altitude polar orbit from September 15, 2009, until moving to a fuel-conserving, 1800 km semi-major axis, elliptical orbit on December 11, 2011 (\citet{Chin2007}). After this date, the change of LRO orbit led to the loss of much data from the spacecraft while LRO was at higher altitudes, limiting the data used for this study (\citet{Barker2021}). One of the LRO mission objectives is to find potential lunar landing sites for crewed missions. The mission also searches for potential resources and characterises the radiation environment. LRO directly produced topographic maps and stereo images from onboard instruments, which can be used to derive digital terrain maps (DTMs). Images taken by the Wide Angle Camera (WAC), part of the LRO Camera (LROC) experiment, were used to derive a global topographic map of the Moon at a 100 m image resolution (\citet{Robinson2010}). LRO is still operating. The Lunar Orbiter Laser Altimeter (LOLA) (\citet{Smith2010}) is one of the six scientific instruments on LRO. LOLA measures topographic elevations to metre elevation accuracy but is limited to 5 m spatial resolution in the best cases and, more often, to 57 m.

\subsection{The Lunar Orbiter Laser Altimeter (LOLA)}
LOLA provides one of the highest-resolution global surface elevation data sets for the Moon. LOLA uses LIDAR (\citet{Smith1997}) to do so. LIDAR measures the round trip Time Of Flight (TOF) for a laser shot reflected off the lunar surface. The LOLA LIDAR consists of a laser transmitter (Nd: YAG in the LOLA case), emitting a wavelength of 1064 nm and a receiver (an aluminium Cassegrain telescope). TOF halved gives the distance from LOLA to the lunar surface from the light travel time, in conjunction with the spacecraft’s well-determined orbit. A similar instrument package was used onboard the earlier Clementine spacecraft (\citet{NRC1997}). LOLA measured the slope, roughness and the 1064 nm reflectance of almost the entire lunar surface on scales of 500 m, 200 m, 50 m and 5 m.

\begin{figure}
    \centering
    \includegraphics[width=\columnwidth]{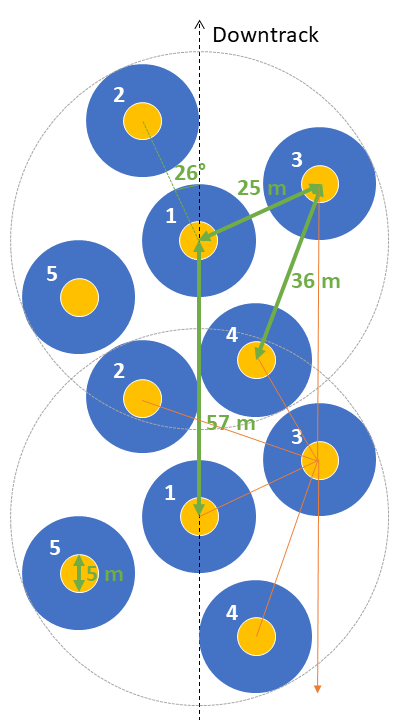}
    \caption{Two LOLA shots encircled in dashed grey with five labelled spots as blue circles. The pattern angle is at a 26° angle to the down-track direction. The distance between shots is 57 m, and the smallest spot-to-spot length is 25 m. Yellow circles display the five illuminated 5 m diameter spot footprints, and blue circles show the field of view of each detector. Orange lines show the six possible slope calculations between spot 3 and the surrounding spots.}
    \label{fig:LOLA_spots}
\end{figure}

LOLA operates by propagating a single laser pulse (laser 1 and 2 alternate monthly) through a Diffractive Optical Element (DOE), splitting the pulse into five separate beams, and rotating 26° to the down-track direction. LOLA’s five pulse spot pattern is illustrated in Figure \ref{fig:LOLA_spots}. LOLA emits short pulses at the rate of 28 Hz (considering the five beams, this is 140 measurements per second) from two lasers (of energies 2.7 mJ and 3.2 mJ for laser 1 and laser 2, respectively). These five beams illuminate the lunar surface in a cross pattern in the far field. Each beam has a diameter of 5 m at the surface. Each beam backscatters off a spot on the lunar surface (yellow in Figure \ref{fig:LOLA_spots}). The receiver telescope detects the returning pulses, and the five-spot pattern is imaged. The laser has a pulse width of $< 10$ ns, and the width of each returned pulse is recorded, measuring the height variation (roughness) within the 5 m footprint of the laser on the surface. Consecutive five-spot patterns are displaced by 57 m due to the orbital motion of LRO. The detector's field of view is shown in blue. The spot pattern calculates slope and surface roughness along a range of azimuths. 

\subsubsection{LOLA Pulse Width Measure}
LOLA also directly measures the roughness within the 5 m footprint of a single spot using the measured pulse width (Section \ref{sec:LOLA}). However, despite the LOLA data being extensively filtered, by the selection of pulses only above 0.15 fJ energy levels and recorded at less than 5° off-nadir angle, the pulse width roughness maps suffered from erroneous values (\citet{Glaser2014}). Pulse width roughness maps will not contribute to site assessment.

\subsection{LOLA Data and Corrections}
NASA’s Planetary Data System (PDS) is a public online archive system where laboratory results, planetary missions and observational data are stored in common descriptions. We used the calibrated, geolocated and aggregated, time-ordered Reduced Data Record (RDR) files (\citet{Neumann2009}) to create LOLA DTMs for the lunar sites of interest. 
LOLA raw data includes two instrumental signatures; their removal is required to make accurate elevation maps. The two signatures are that (1) the LIDAR pulses are not symmetric, and (2) there are biases in the TOF calculation. These biases vary for each of the five detectors. Both effects have been modelled by \citet{Riris2009}. They find (1) pulse asymmetry: a variable bias dependent on the received pulse strength arises from an induced range bias and asymmetry of the detector’s electronic impulse response leading to a distortion of the received pulse. (2) A TOF bias: each channel has different cable lengths, leading to a fixed offset. The fixed and variable offset corrections are applied to the pulse width datasets; the fixed offsets can be found in \citet{Riris2009}.

The receiver pulse width error calculated by standard deviation remains below 800 ps overall energies, equivalent to 0.24 m in elevation. From 10 ns to 25 ns, the received pulse widths have residuals compared with the emitted pulse width at all pulse widths of $\approx 1$ ns, equivalent to 0.3 m elevation error (\citet{Riris2009}). The residuals are larger for pulse widths $< 10$ ns, $\approx 5$ ns, 1.5 m.

The co-registration techniques of \citet{Glaser2013} were applied to the individual LOLA tracks that are incorporated in the DTMs. Co-/Self-registration is an algorithm identify misaligned measurements and tracks from the inputs of LOLA data and DTMs to which laser altimeter profiles are registered (\citet{Scholten2012, Scholten2011, Glaser2017}). A 0.13 m to several meters (depending on the number of measured laser shots) positional accuracy and 0.18 m residual heights can be achieved for data sets. This can be compared with the pre-registration accuracy of the LOLA data, \citet{Araki2009, Bussey2003} found positional errors of $\sim$4 m radially and $\sim$77 m horizontally. There are negligible for the $\sim 200$ m scale being probed here.

\subsection{Data Loss from a Thermal Anomaly}
\label{sec:anomaly}
There is a thermally-induced anomaly in the LOLA instrumentation. All five channels work nominally when the spacecraft is over the lunar day side; however, only two of the five receiver channels acquire significant data when LRO observes during the cold lunar night. This ‘thermal anomaly’ was found in ground testing due to the thermal blanket's contraction in the cold. This contraction pulls the transmitter beam out of focus with the receiver (\citet{Smith2017}). Detectors 1, 2 and 5 in Figure \ref{fig:LOLA_spots} do not operate on the ‘night side’. Detectors 3 and 4 align to emitted spots 2 and 5 by a fortunate coincidence, enabling continuous observation (\citet{Glaser2014}).

For instance, the effects of the LOLA thermal anomaly can be observed in the Mare Moscoviense region. The variation in the number of spots detected per shot means LOLA is less effective for this site. Only $47\%$ of shots yield detections in all five spots; another $41\%$ of the area is illuminated by two spots per shot; $9\%$ is illuminated by a single spot per shot; $2\%$ is covered by four spots per shot; $1\%$ is covered by three spots per shot. The impact of this anomaly is important when constructing slope and roughness maps because there are inconsistencies between the number of spots per fitted plane (Section \ref{sec:Method}), which limits the spatial resolution of these maps.

\begin{figure*}
    \centering
    \begin{subfigure}[t]{0.3\textwidth}
        \centering
        \includegraphics[width=\textwidth]{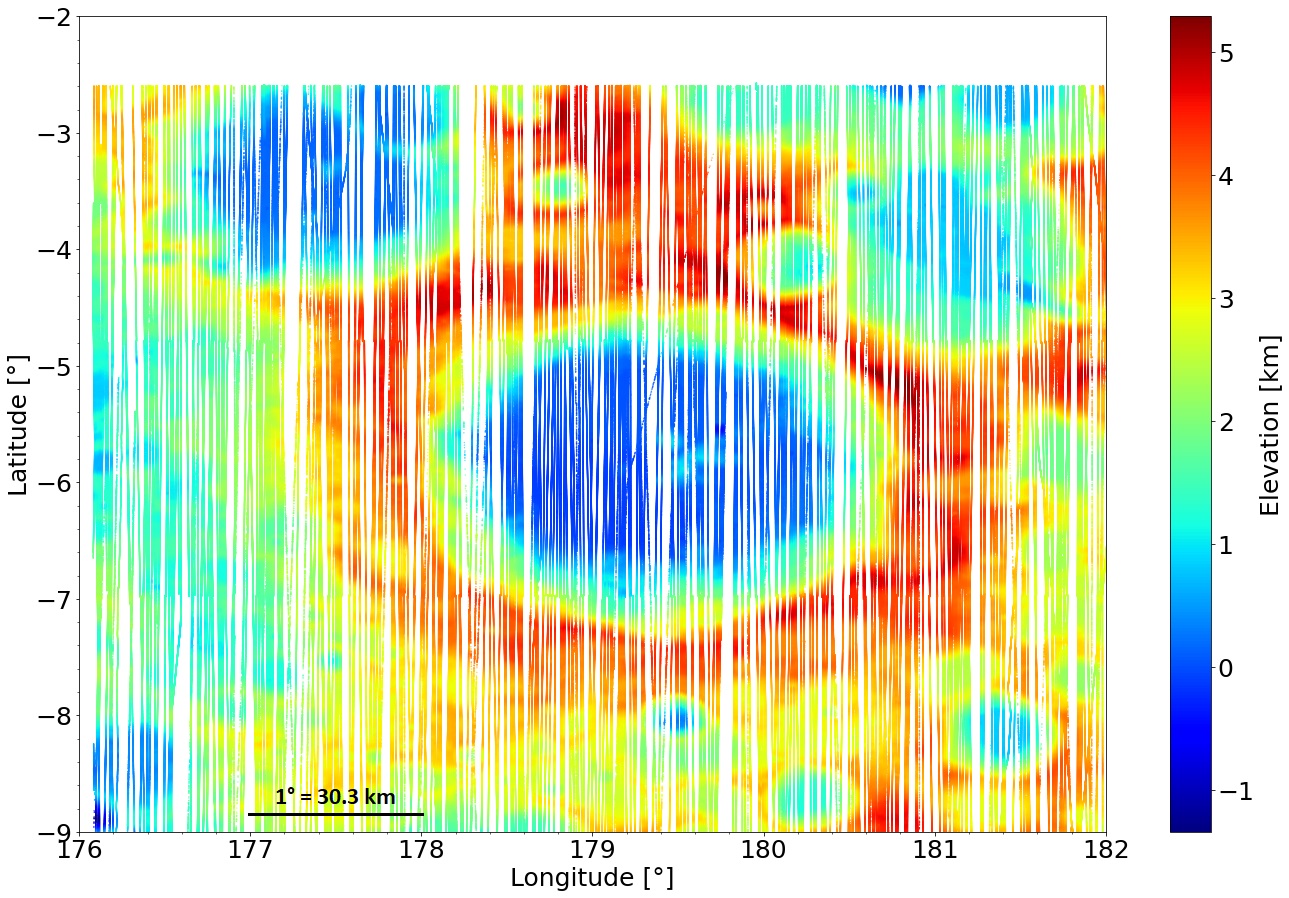}
        \caption{}\label{fig:non_interp}
    \end{subfigure}
    \hfill
    \begin{subfigure}[t]{0.3\textwidth}
        \centering
        \includegraphics[width=\textwidth]{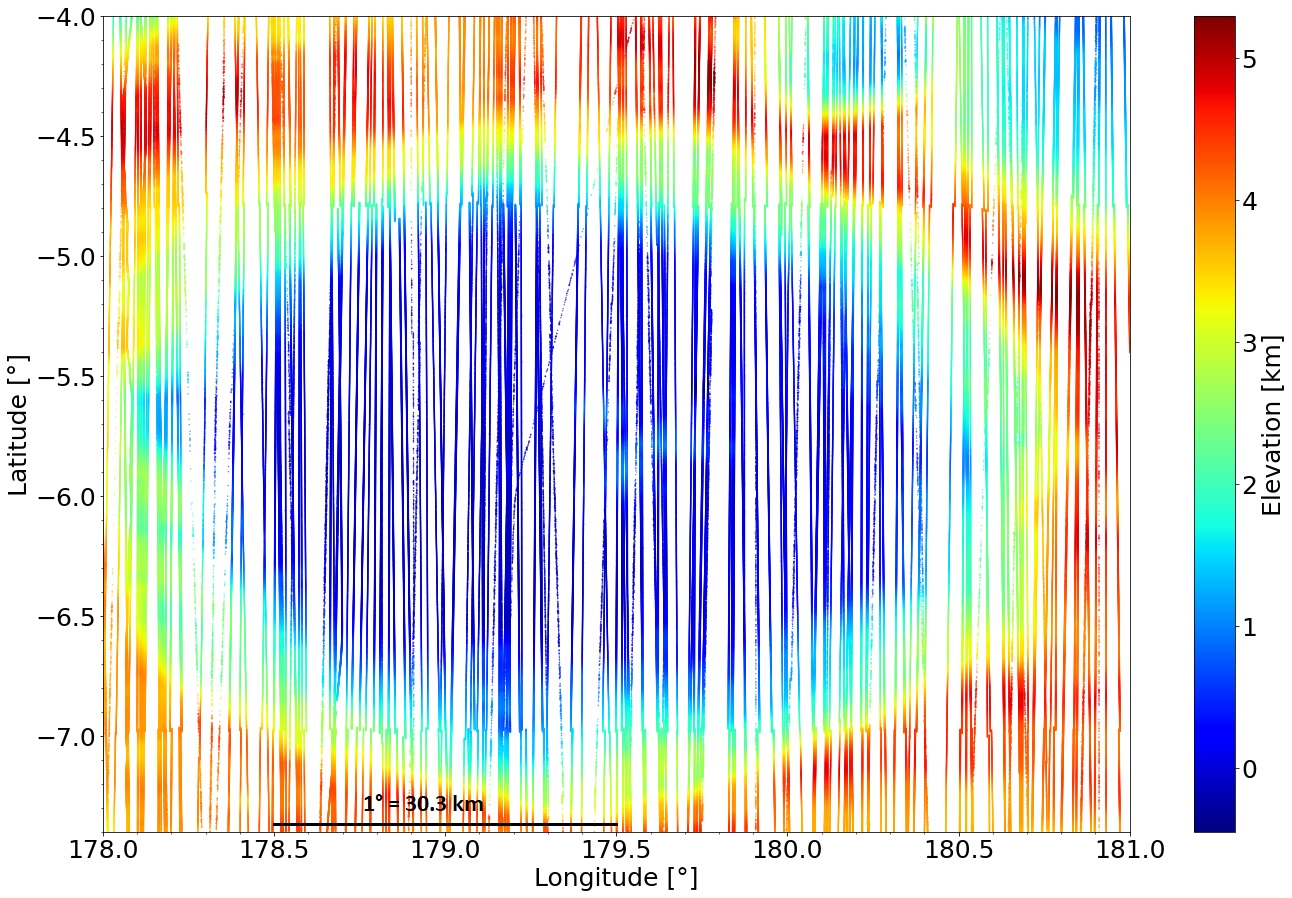}
        \caption{}\label{fig:small_non_interp}
    \end{subfigure}
    \hfill
    \begin{subfigure}[t]{0.3\textwidth}
        \centering
        \includegraphics[width=\textwidth]{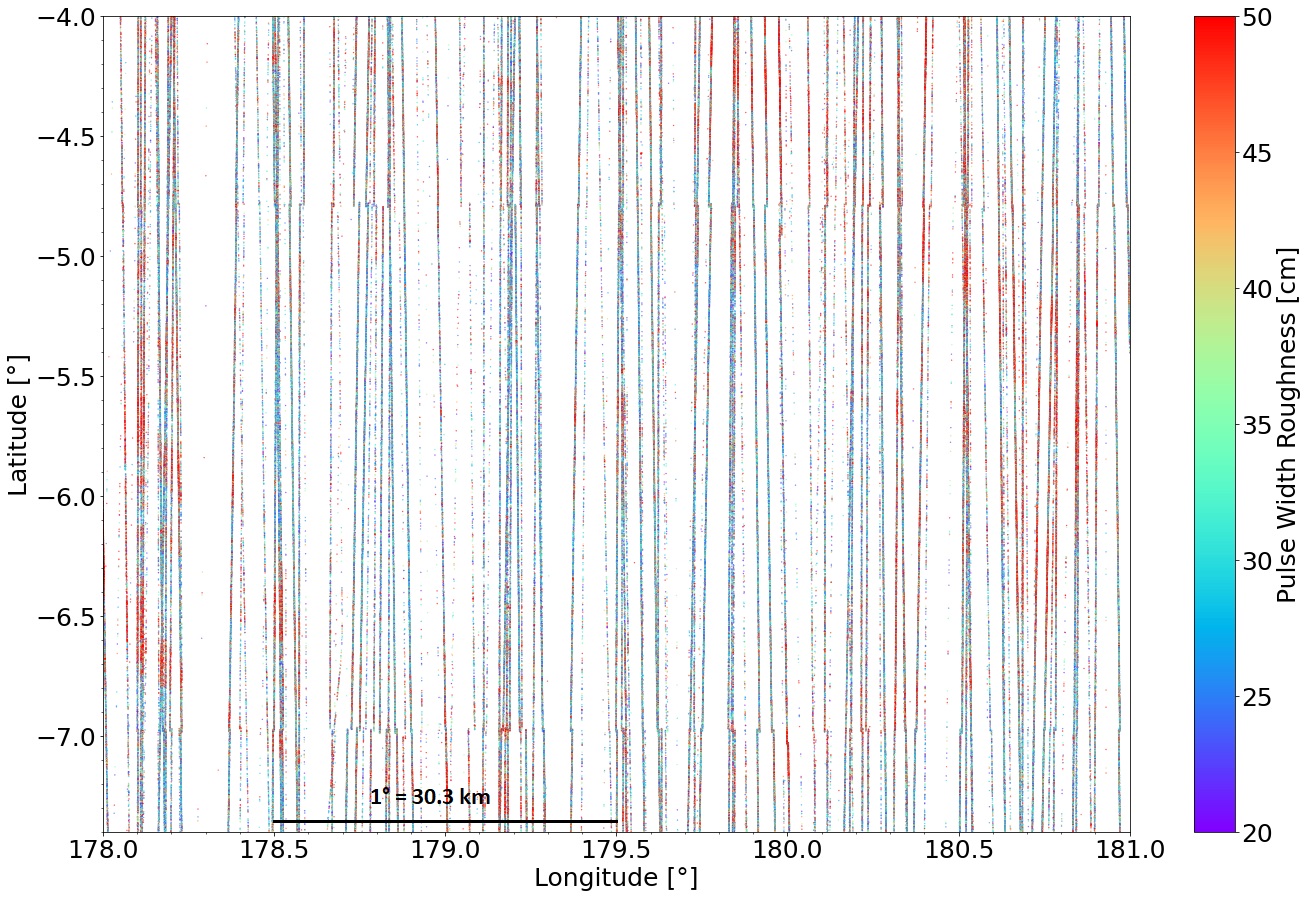}
        \caption{}\label{fig:non_interp_pw}
    \end{subfigure}
    \caption{Topographic maps of the lunar crater Daedalus. (a) A DEM of a 181 x 196 km region where a colour scale represents the elevation in km. Data points from the 5 m footprints are coloured and white means there is no data at those locations. (b) A zoomed-in DEM of a 91 x 106 km region where a colour scale represents the elevation in km. (c) A zoomed-in pulse width roughness map of a 91 x 106 km region where a colour scale represents the roughness in cm. The pulse width data is measured by broadening the received pulse to the transmitted pulse on the 5 m footprint scale.}\label{fig:non_interpolated}
\end{figure*}

The smallest lunar site used in this study is Daedalus. Figure \ref{fig:non_interpolated} shows LOLA track coverage maps of the smaller site for elevation and pulse width roughness data. The larger region (Figure \ref{fig:non_interp}), $181 x 196$ km in size, consists of 4,614,796 data points. When zooming in on the site to a $91 x 106$ km region (Figure \ref{fig:small_non_interp}) the number of data points decreases to 1,138,030 and is covered by 203 LOLA tracks. The small $91 x 106$ km region measured 431,842 pulse width roughness points and is covered by 179 LOLA tracks. The coverage maps show that the centre of Daedalus has data gaps with maximum horizontal separation $\approx 6$ km. As a result, the pulse width roughness map has worse coverage and alternative methods to derive roughness are discussed in Section \ref{sec:RMS}.

\section{Candidate Site Selection}
\label{sec:candidate}
\begin{figure}
    \centering
    \includegraphics[width=\columnwidth]{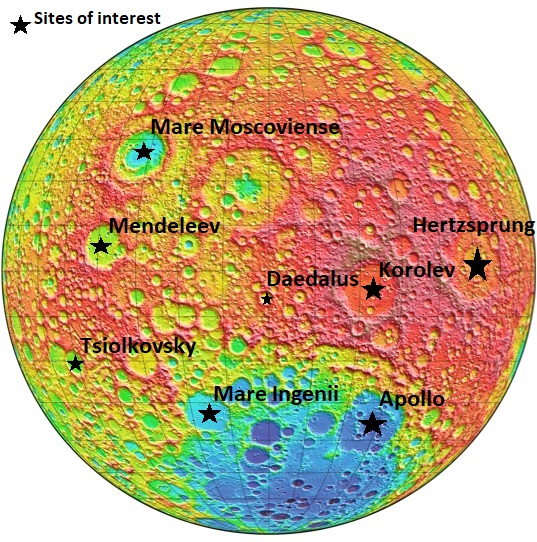}
    \caption{Topographic map of surface height on the lunar farside using LOLA data (image available on NASA's PDS \citep{NASA}). The data shows the mountainous terrain and distribution of impact craters. The sites of interest assessed in this paper are marked with black stars and labelled.}
    \label{fig:farside}
\end{figure}

\begin{figure*}
    \centering
    \includegraphics[width=\textwidth]{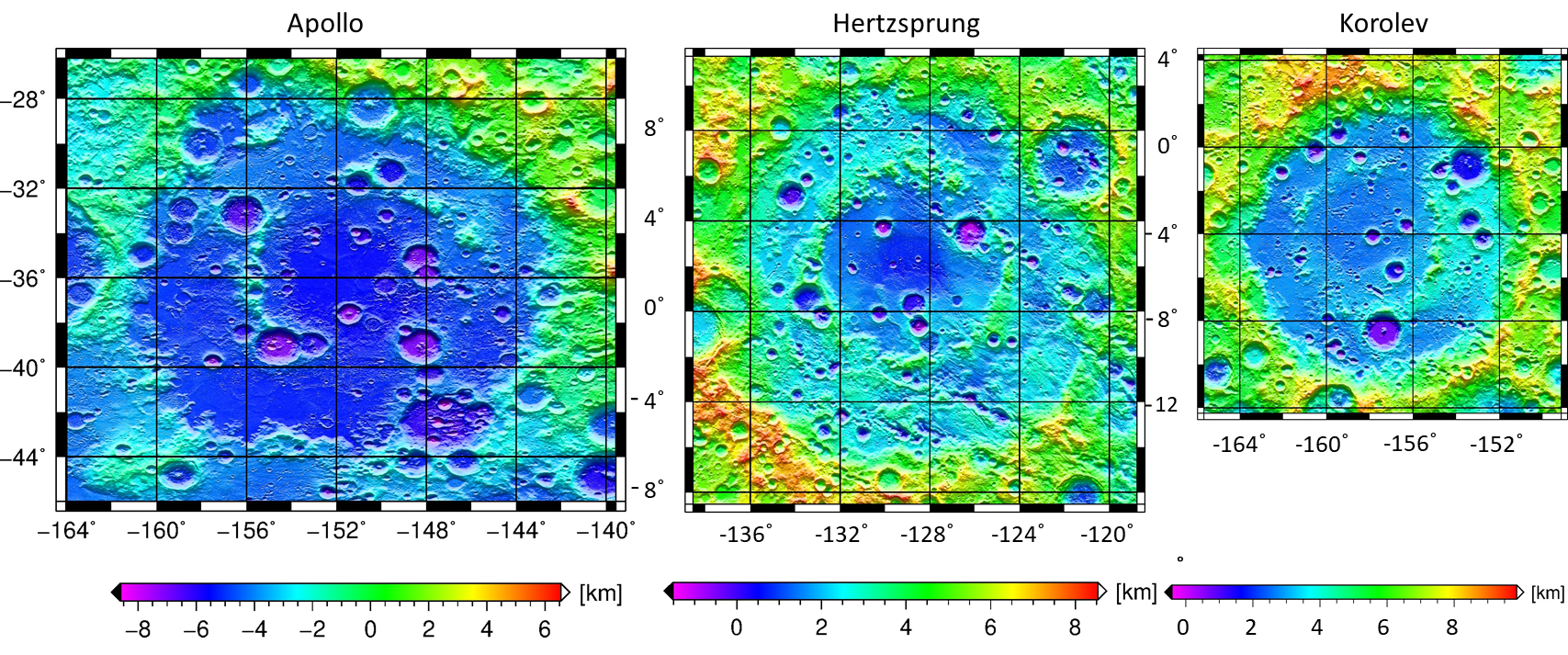}
    \includegraphics[width=\textwidth]{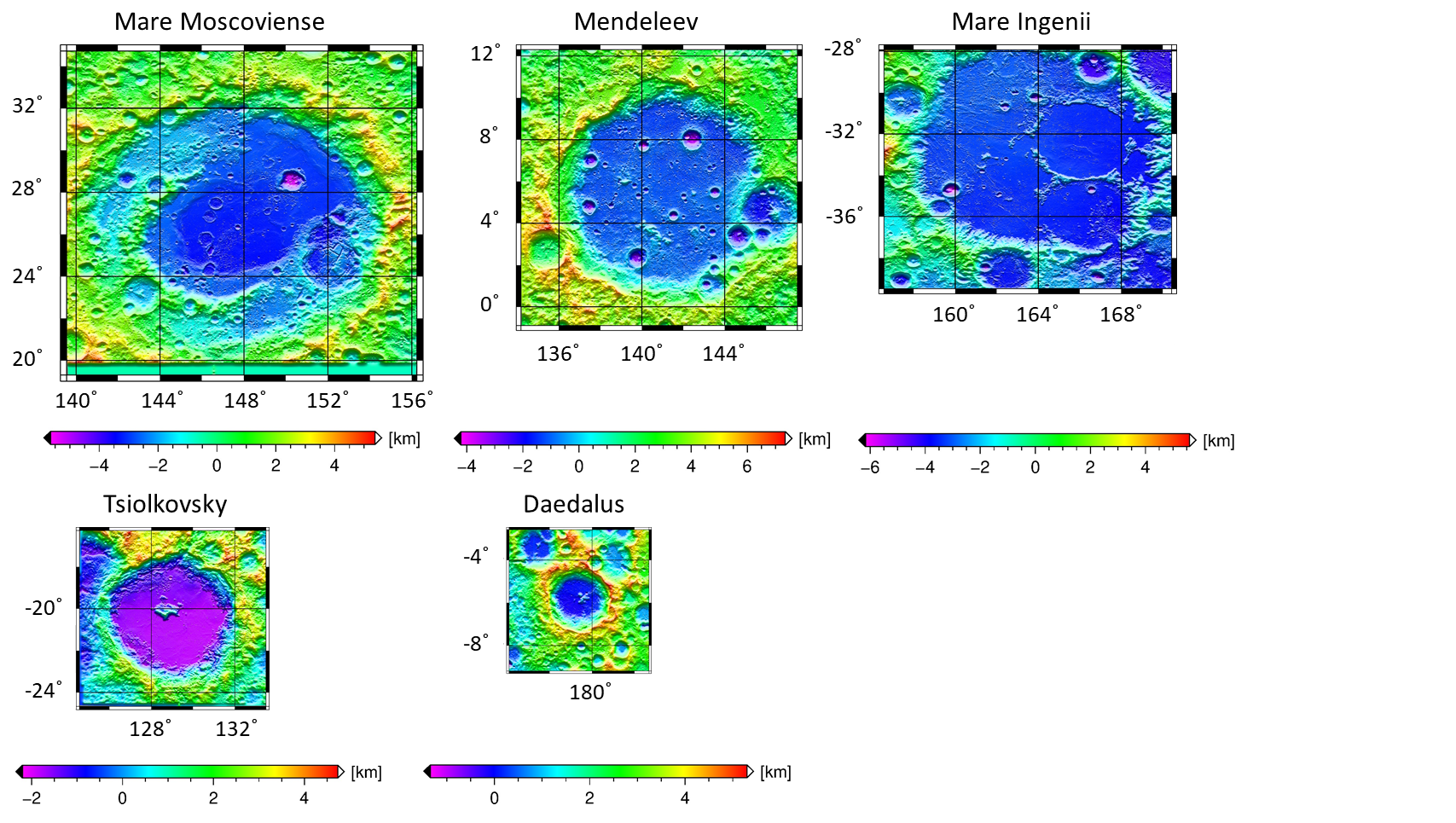}
    \includegraphics[width=\textwidth]{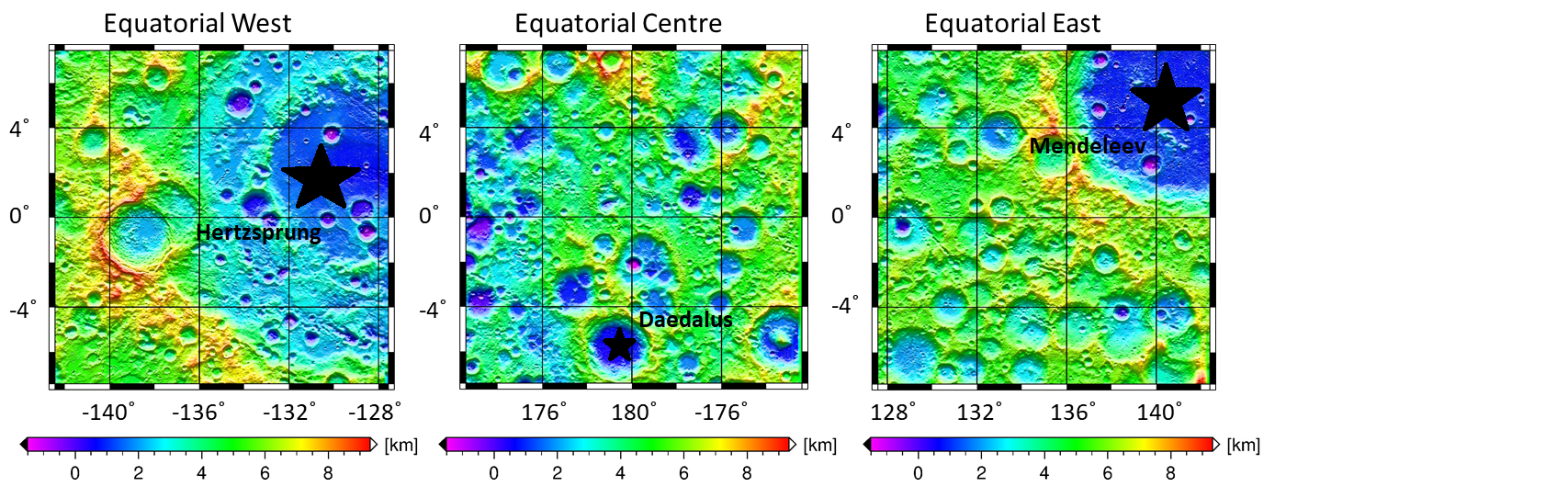}
    \caption{Interpolated digital elevation maps of the eight sites of interest and three equatorial sites, where a pixel is 200 m long. Elevation in km is shown by a colour scale below each map and the site name is above each map. The maps are overlaid by a 4° in length grid ($\sim$ 122 km).}\label{fig:sizes}
\end{figure*}

The search for a suitable site began with the visual selection of eight large ($> 100$ km diameter) maria\footnote{Maria are identified as large darker surfaces formed by mare volcanism (\citet{Shultz1983}).} and craters within the RQZ (longitude $> 108$°) because they have a higher probability of having a smooth crater floor. The locations of the eight lunar far-side sites are shown in Figure \ref{fig:farside}. Their properties are listed in Table \ref{tab:sites}. The third column of Table \ref{tab:sites} gives the distance in degrees of the lunar sites past the RQZ boundary. Elevation maps of the eight sites of interest are shown in Figure \ref{fig:sizes} to give a visual impression of their topography. A 4°$x$4° ($122 x 122$ km) grid is overlaid on each site map, showing their relative sizes.

In addition, three comparison equatorial regions were studied on the lunar far-side. At longitudes +30°, 0° and -30° from the anti-Earth point. These equatorial regions are evidently rougher terrain than the eight candidate sites (Figure \ref{fig:sizes}).

\begin{table*} 
\caption{A table of the chosen sites and their properties, labelled by name. Their central coordinates are listed in the Selenographic coordinate system. Degrees into the Radio Quiet Zone (RQZ) is the distance in degrees past 147° East or West. The diameter of each site is the longest horizontal length within the crater boundaries. The minimum and maximum elevation is determined from LOLA data and is in reference to the Moon's radius, 1,736 km. Below the eight chosen sites are three equatorial positions mapped to show how mountainous the far-side is. \label{tab:sites}}
    \begin{tabular}{lccccc}
        \hline
        \multicolumn{1}{l}{Site} & \multicolumn{1}{l}{Coordinates} & \multicolumn{1}{l}{Degrees into RQZ} & \multicolumn{1}{l}{Diameter (km)} & \multicolumn{1}{l}{Min Elevation (km)} & \multicolumn{1}{l}{Max Elevation (km)} 
        \\
        \hline
        \multicolumn{6}{l}{Candidate Far-Side Radio Array Sites} \\
        \hline
        Apollo & 36.1°S 151.8°W & 49.6°W & 537 & -8.6 & -1.4
        \\
        Daedalus & 5.9°S 179.4°E & 77.2°E & 93 & -0.4 & 4.1
        \\
        Hertzsprung & 1.4°N 128.7°W & 26.5°W & 570 & -1.5 & 5.3 
        \\
        Korolev & 4.0°S 157.4°W & 55.2°W & 437 & -0.1 & 8.7 
        \\
        Mare Ingenii & 33.7°S 163.5°E & 61.3°E & 282 & -5.8 & 0.6
        \\
        Mare Moscoviense & 27.3°N 147.9°E & 45.7°E & 276  & -5.6 & 4.2 
        \\
        Mendeleev & 5.7°N 140.9°E & 38.7°E & 313 & -4.2 & 5.8 
        \\
        Tsiolkovsky & 20.4°S 129.1°E & 26.9°E & 180 & -2.2 & 4.2 
        \\
        \hline
        \multicolumn{6}{l}{Comparative Equatorial Regions}
        \\
        \hline
        Equatorial Central& 0°N 180°E & 77.8°E & 445 & -1.6 & 6.6
        \\
        Equatorial East & 0°N 135°E & 32.8°E & 445 & -3.6 & 7.5 
        \\
        Equatorial West & 0°N 135°W & 32.8°W & 445 & -1.5 & 9.4\\
        \hline
    \end{tabular}
\end{table*}

\section{Data Analysis Methods}
\label{sec:Method}
\subsection{Digital Elevation Maps (DEM)}
\label{sec:DEM}
More quantitative topographic information is derived from constructing Digital Elevation Maps (DEMs) from the LOLA data. DEMs display surface topography by mapping the height variation of a region. DEMs were generated by the Generic Mapping Tools software (GMT) (\citet{Wessel2019}). GMT is a free and open source code\footnote{https://docs.generic-mapping-tools.org/6.4/} that allows for manipulating data parameters to produce sophisticated illustrations for Earth, ocean and planetary sciences. The motivation for creating DEMs is to determine whether a site has approximately 200 km regions of smooth, low-slope terrain and to note any physical barriers, such as significant mountainous or crater features. 

The LOLA data sets (Longitude, Latitude, Elevation) are projected onto a grid and interpolated to create a regular grid. Given a discrete data set, interpolation derives a polynomial function which passes through the provided data, enabling intermediate values to be estimated. The generated grid files can be read into Python and MATLAB to produce elevation maps. These tools enable 3D interactive elevation created to be used to visualise these sites from differing and exaggerated perspectives.

Non-interpolated DEMs were also constructed, where elevation values are assigned to a colour scale and plotted on a longitude-latitude plane. These images are useful as they highlight the non-uniform track coverage of LOLA. These maps are also convenient for laser spot detection analysis where the LOLA thermal anomaly can be observed for each site (Section \ref{sec:anomaly}). The DEMs can then be used to generate slope and roughness maps.

\subsection{Slope}
The slope is defined as the terrain height variance over a specified distance and so is a function of the distance over which the slope is measured, the length scale. Several measures of surface roughness have been defined in literature \citep[e.g.,][]{Kreslavsky2000,Shepard2001}. These include both one-dimensional and two-dimensional slopes.

\subsubsection{One-Dimensional Slope}
Discussed here are three one-dimensional methods to derive slopes from LOLA data: (1) RMS slope; (2) median absolute slope; (3) median differential slope for a range of lengths. 

(1) The RMS slope is defined in one-dimension as the Root Mean Square (RMS) difference in height, $\Delta z$ between two points (also called the deviation, $v$) over the distance between the points, $\Delta x$:
\begin{equation}
    s(\Delta x) = \frac{\Delta v(x)}{\Delta x} = \frac{\langle [z(x_{i}) - z(x_{i-1})]^{2} \rangle ^{\frac{1}{2}}}{\Delta x}
\end{equation}
where angular brackets indicate the mean of the bracket contents (\citet{Rosenburg2011}).

(2) The median absolute slope can be derived on the smallest scales, $\sim25$ m, the slope of the central spot and one of the four edge spots, shown in Figure \ref{fig:LOLA_spots} (\citet{Rosenburg2011}). 

(3) The median differential slope (\citet{Kreslavsky2000}) is derived as follows. For a given baseline, $L$, through five points, find the difference in elevation for a point half the baseline ahead, $z_{\frac{1}{2}}$, and a point half the baseline behind, $z_{-\frac{1}{2}}$. Calculate the elevation difference for points a baseline ahead, $z_{1}$, and a baseline behind, $z_{-1}$. Subtract half the latter derived elevation difference from the first elevation difference. 
\begin{equation}
    \tan(\alpha) = \frac{\left(z_{\frac{1}{2}} - z_{-\frac{1}{2}}\right) - \frac{1}{2}\left(z_{1} - z_{-1}\right)}{L}
\end{equation}
The resultant elevation difference over the baseline is the tangential slope, $\alpha$. 

Each method has strengths and weaknesses: (1) RMS is an established method because it is also used to measure the scatter of radar reflection. However, the method is sensitive to outliers because of its dependence on the deviation squared. (1,2) The RMS and median absolute slope are one-dimensional slope methods derived in the down-track direction. Therefore, both underestimate the surface gradient if the steepest slope diverges from this direction. (3) Median differential slope removes small-scale and large-scale surface roughness features. Arguably, the median differential slope can be described as an intuitive parameter because small-scale roughness is measured with respect to the long wavelength roughness profile. The median differential slope method is a better measure compared to the RMS slope or median absolute slope method because natural surface slope-frequency distributions are commonly non-Gaussian with long tails.

\subsubsection{Two-Dimensional Slope}
\label{sec:Slope}
A two-dimensional slope is preferred because slopes that diverge from the down-track direction are included. The two-dimensional slope can be derived from multiple spot points within a LOLA shot, e.g., between spots 1-3-4 in Figure \ref{fig:LOLA_spots}. A total of six slope measurements can be derived from spot 3 in Figure \ref{fig:LOLA_spots} in the directions shown in orange. Vector geometry computes the plane through three spots, identifying slope magnitude and the azimuth of the slope. The baseline in this method is determined as the square root of the area of the triangle. As discussed in Section \ref{sec:anomaly}, LOLA experiences a thermal anomaly which reduces the number of spots from which signals are detected. An interpolated grid has to be used to provide a consistent distance over which slopes are calculated, and a vectorized method is the most computationally efficient.

\subsection{Roughness}
\label{sec:RMS}
Roughness can be defined as the Root Mean Square (RMS) deviation from a specified plane (\citet{Glaser2014}), i.e., the scatter about a one- or two-dimensional slope. Roughness, similarly to slope, is dependent on the length scale and is calculated here on a 200 m scale because smaller scales show missing LOLA data tracks. Therefore, slope and roughness cannot be reliably determined. The plane fitting method used in Section \ref{sec:Slope} measures roughness.

The pulse width measurements from LOLA are not used in this study (Section \ref{sec:LOLA}); instead, the $Sigma-Z$ method was used. This method calculates the standard deviation, $\sigma$, of the LOLA elevation, $Z$, spot measurements to a plane. Alternatively, a method known as roving-window analysis can be used, in which a $3x3$ kernel is scanned over the interpolated DEM maps and assigns the deviation in elevation of surrounding pixels to a given pixel. This method is more flexibly applied to various scales.

\section{Map Products}
\label{sec:Results}
\subsection{Equatorial Regions}
\begin{figure*}
    \centering
    \begin{subfigure}[t]{0.3\textwidth}
        \includegraphics[width=\textwidth]{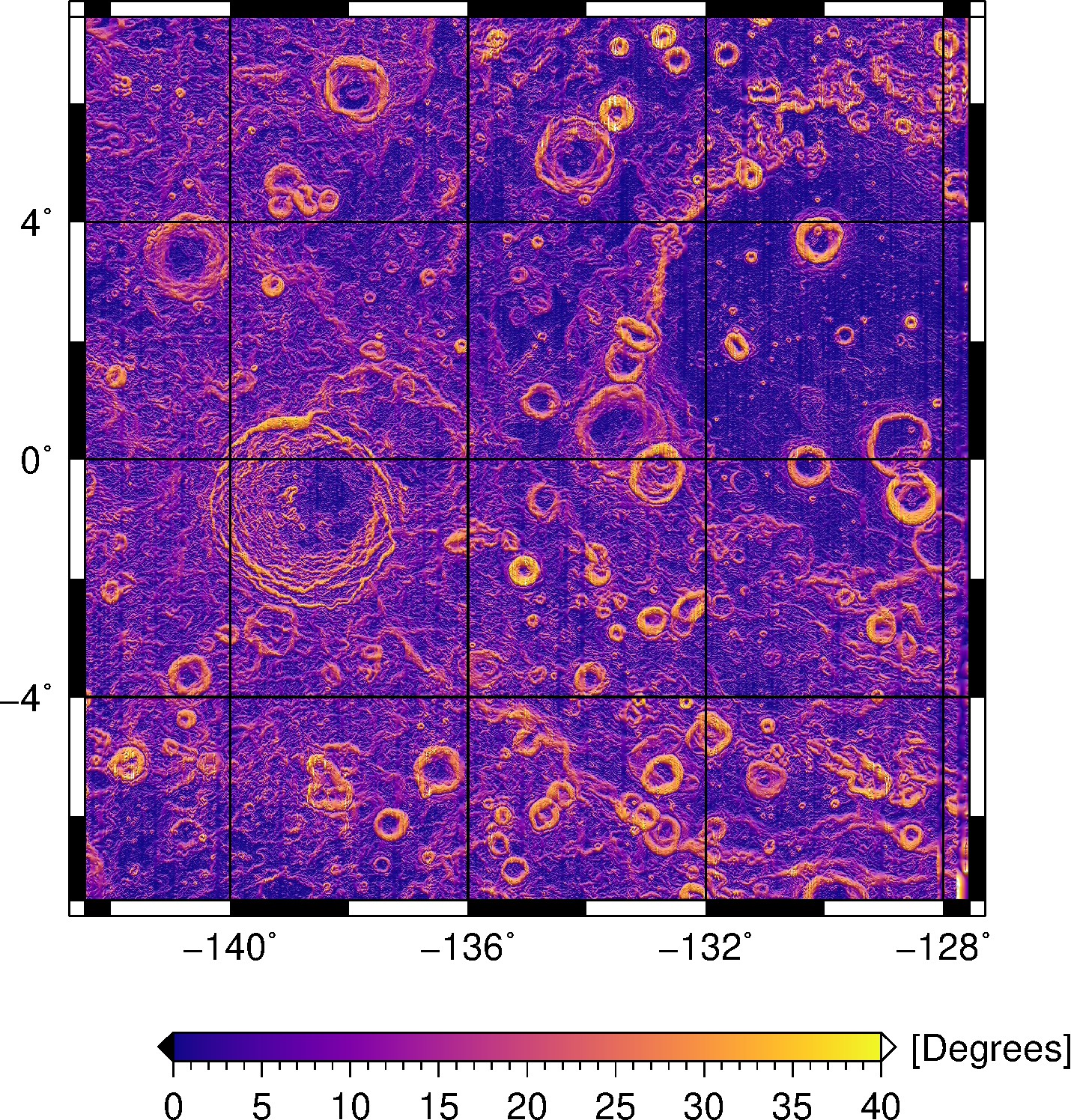}
        \caption{}\label{fig:west}
    \end{subfigure}
    \hfill
    \begin{subfigure}[t]{0.3\textwidth}
        \includegraphics[width=\textwidth]{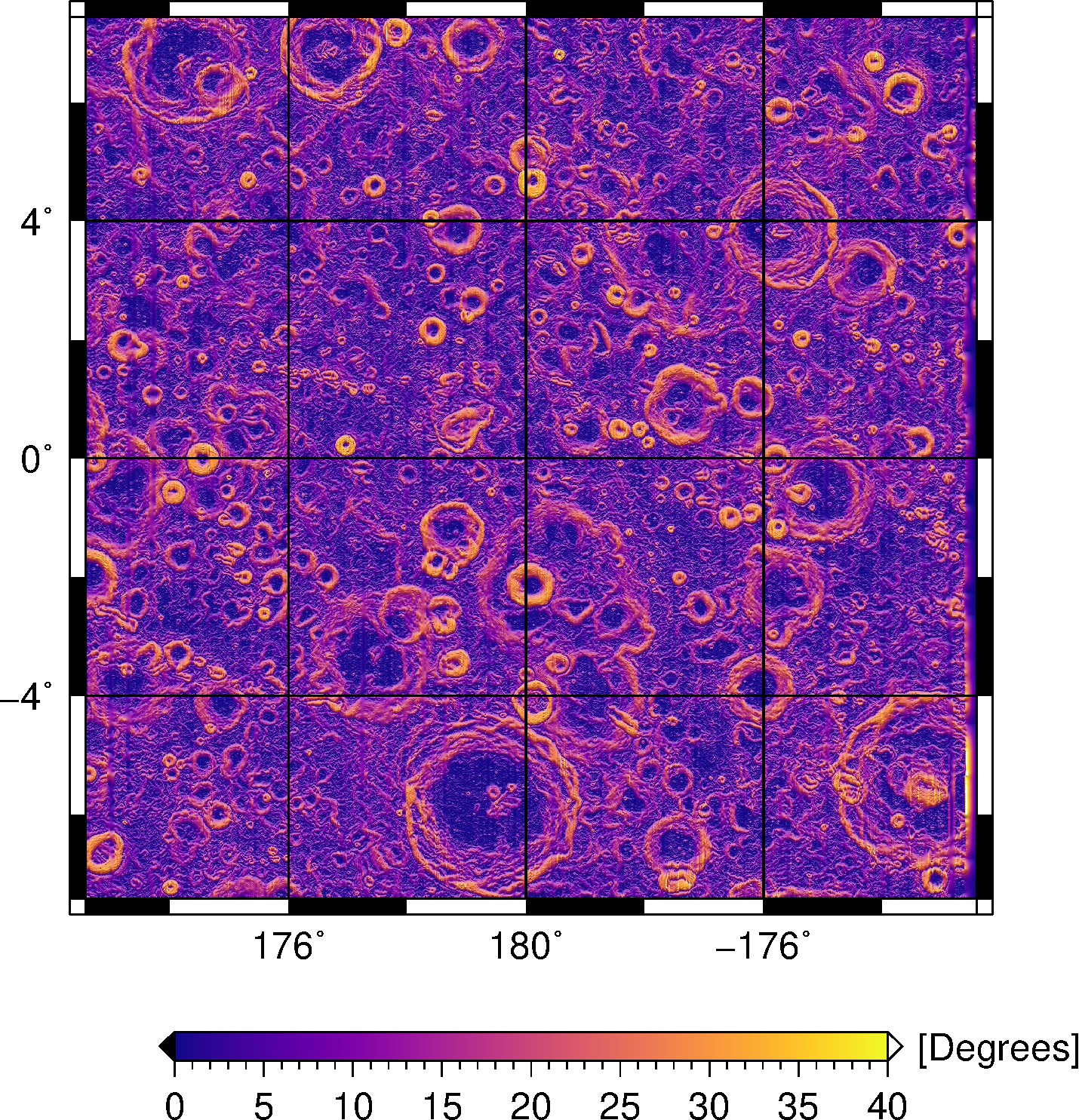}
        \caption{}\label{fig:cent}
    \end{subfigure}
    \hfill
    \begin{subfigure}[t]{0.3\textwidth}
        \includegraphics[width=\textwidth]{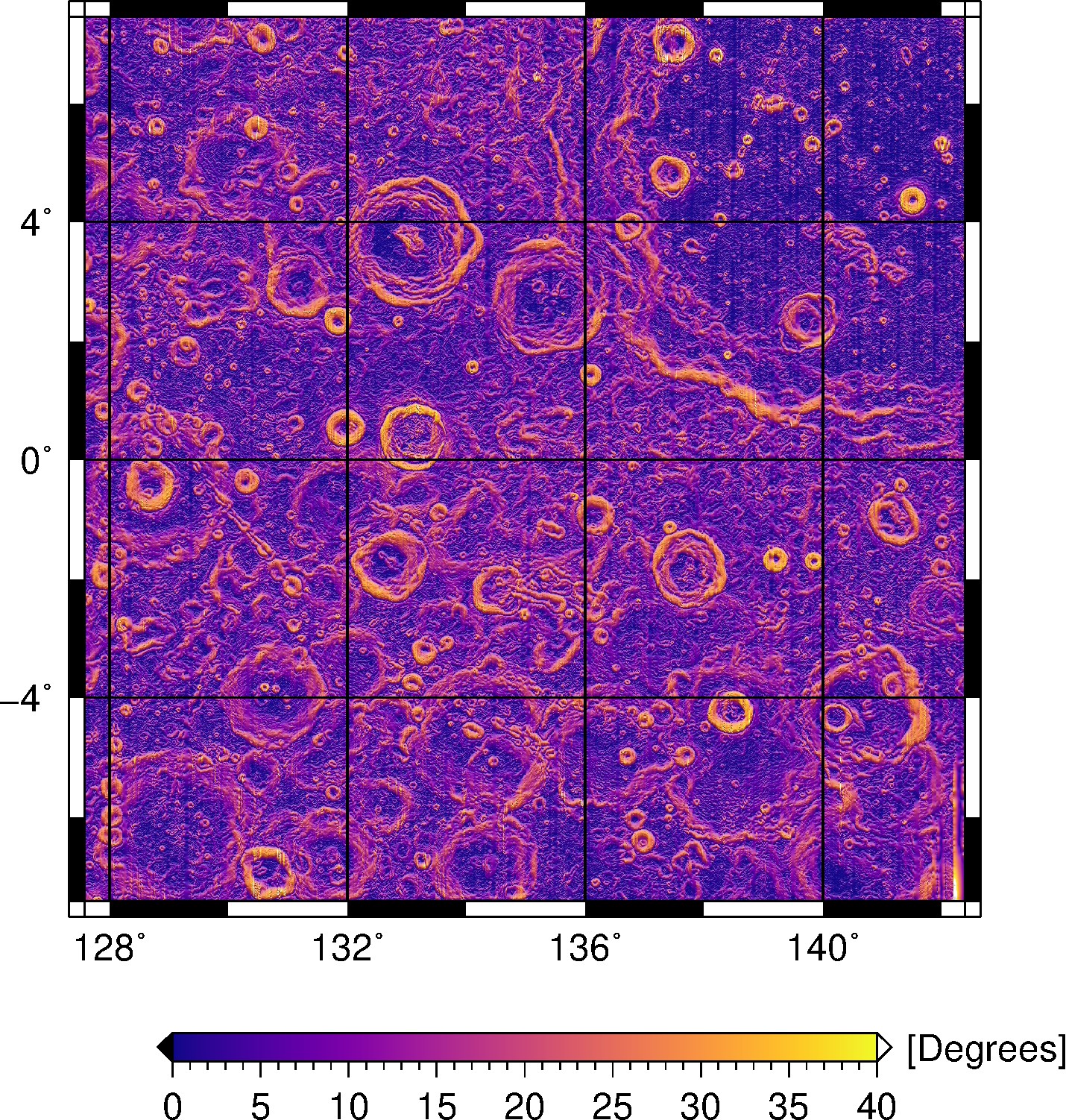}
        \caption{}\label{fig:east}
    \end{subfigure}
    \caption{Slope maps of three 445 km squared regions centred on the lunar far-side equator (Latitude = 0°). The slope is in the degree range of 0-40° over a 200 m scale, represented by a purple colour scale. Dark purple is a smooth surface, and yellow is an inclined surface. (a) The West equatorial region centred at coordinates (-135°,0°). (b) The central equatorial region centred at coordinates (180°,0°). (c) The East equatorial region centred at coordinates (+135°,0°).}
    \label{fig:site_comparison}
\end{figure*}

The three equatorial regions are mapped to show the typical mountainous surface of the lunar far-side. Figure \ref{fig:site_comparison} shows the slope maps for three 445 km squared regions displaced by 45° along the lunar equator. The maps are overlaid by a 4°$x$4° grid ($122 x 122$ km), which shows that within these grids, a traversable surface in the length of hundreds of kilometres is not obtainable. The anti-Earth region (Figure \ref{fig:cent}) shows no traversable path with inclines < 15°, and at inclines < 25°, two perpendicular paths on the order of hundreds of kilometres cannot be achieved. The East and West regions are similar to the central region, but they both show an area of large impact craters which is a smoother surface (dark purple) in Figures \ref{fig:west} and \ref{fig:east}. These areas could be accessible to a rover on inclines < 25° and on the scales of an interferometer. However, the surface is disrupted by many smaller craters with high slopes (< 25°), so deploying a full-sized 200 km array becomes challenging. Instead, sites for baselines a few kilometres in length are quite easy to find.

The highly variable slope of the far-side demonstrates a need for the identification of large smooth marina and craters. 

\subsection{Digital Elevation Maps}
The results for the Mare Moscoviense are described in detail for each map type below. We have chosen to present the results of Mare Moscoviense because the site is a sizable candidate location for an interferometer for its lack of obscuring terrain. The corresponding maps for all eight sites are listed in Table \ref{tab:sites}, and presented the Appendix \ref{app:maps}.

Mare Moscoviense has a mare floor 3 to 4 km below the lunar equatorial radius (1737.4 km, (\citet{Williams2021})), spanning the longest length of 280 km. The mare edge becomes extremely mountainous, reaching up to 4 km above the lunar equatorial radius and forming a steep border to the mare floor. 

\begin{figure*}
    \centering
    \begin{subfigure}[b]{0.3\textwidth}
        \centering
        \includegraphics[width=\textwidth]{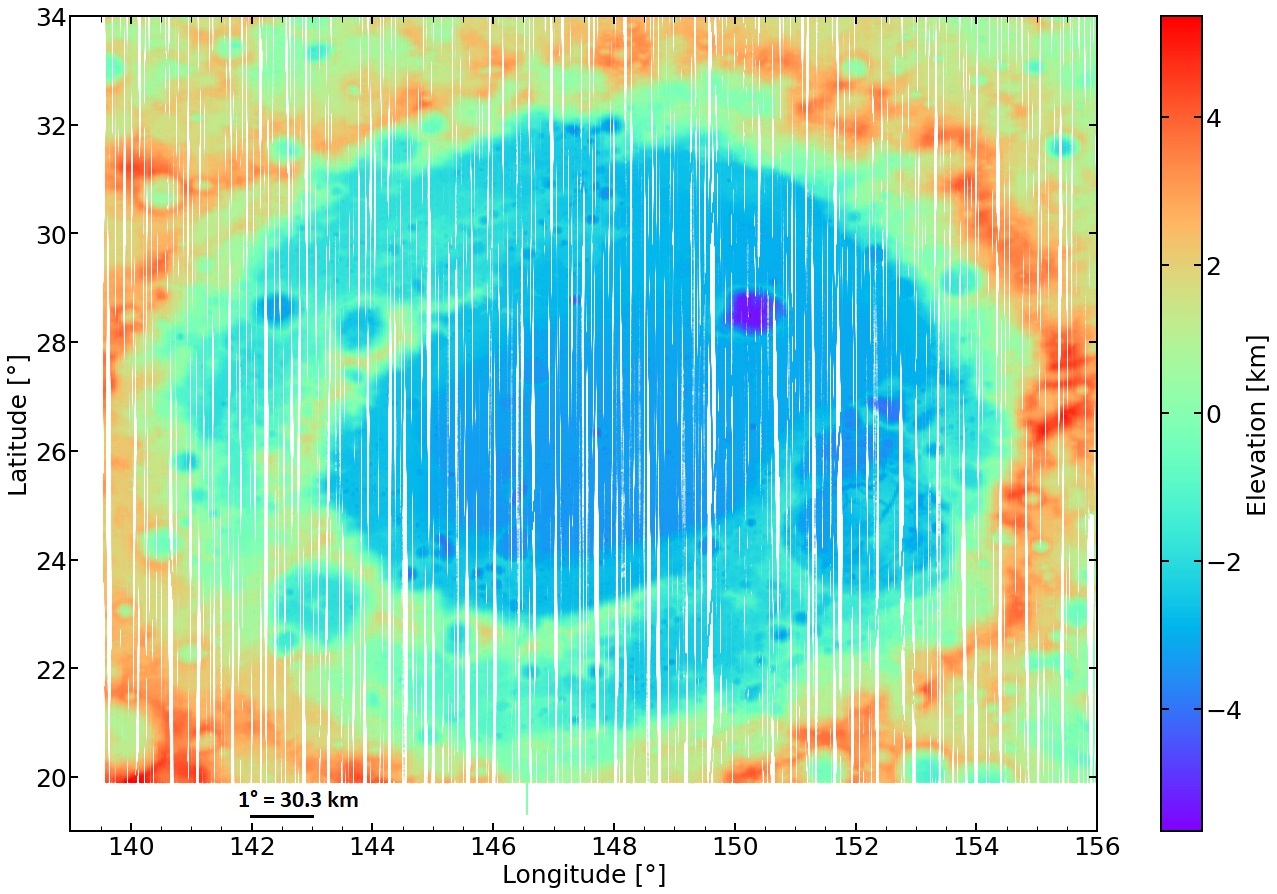}
        \caption{}
    \end{subfigure}
    \begin{subfigure}[b]{0.2\textwidth}
        \centering
        \includegraphics[width=\textwidth]{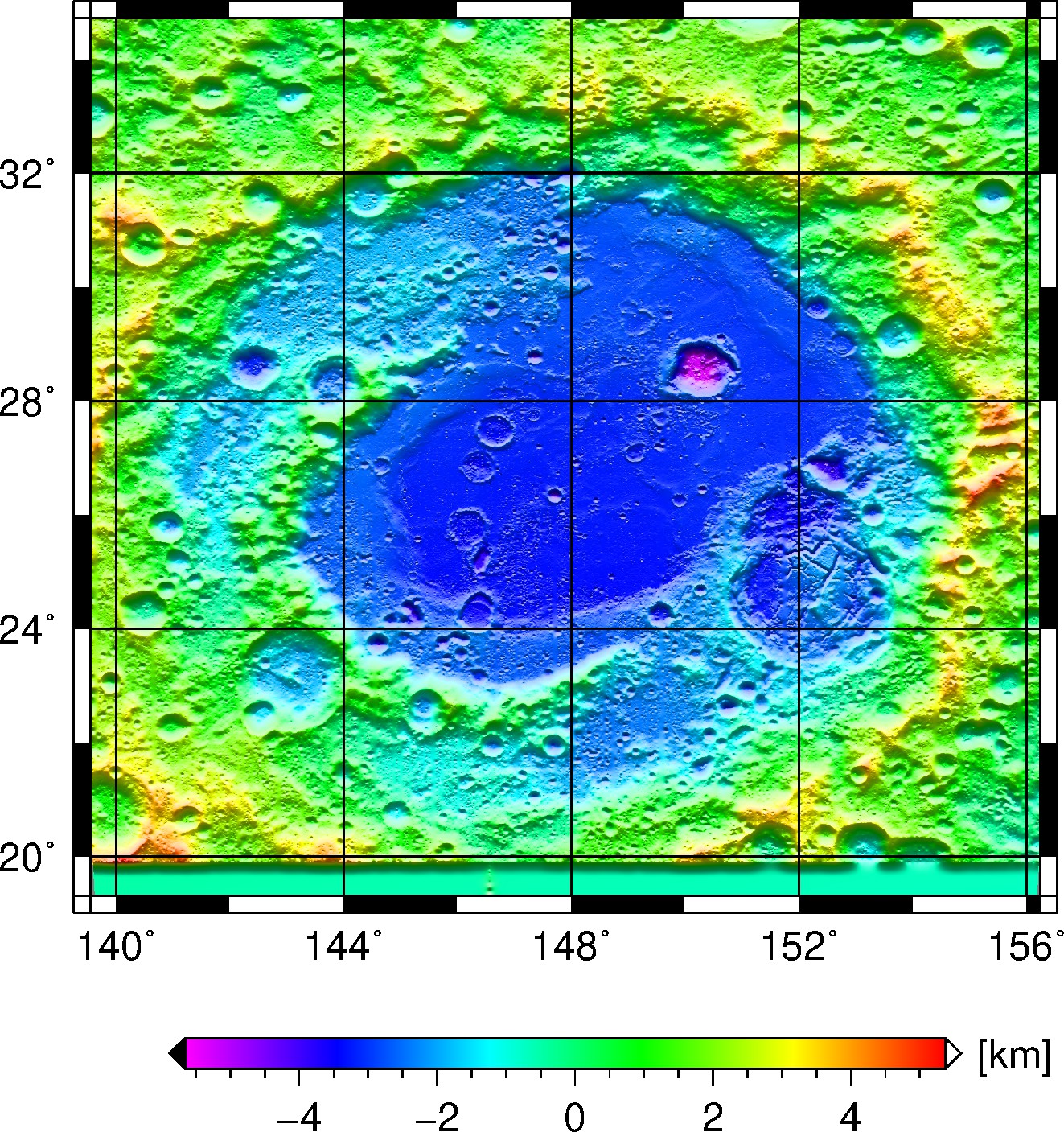}
        \caption{}
    \end{subfigure}
    \begin{subfigure}[b]{0.4\textwidth}
        \centering
        \includegraphics[width=\textwidth]{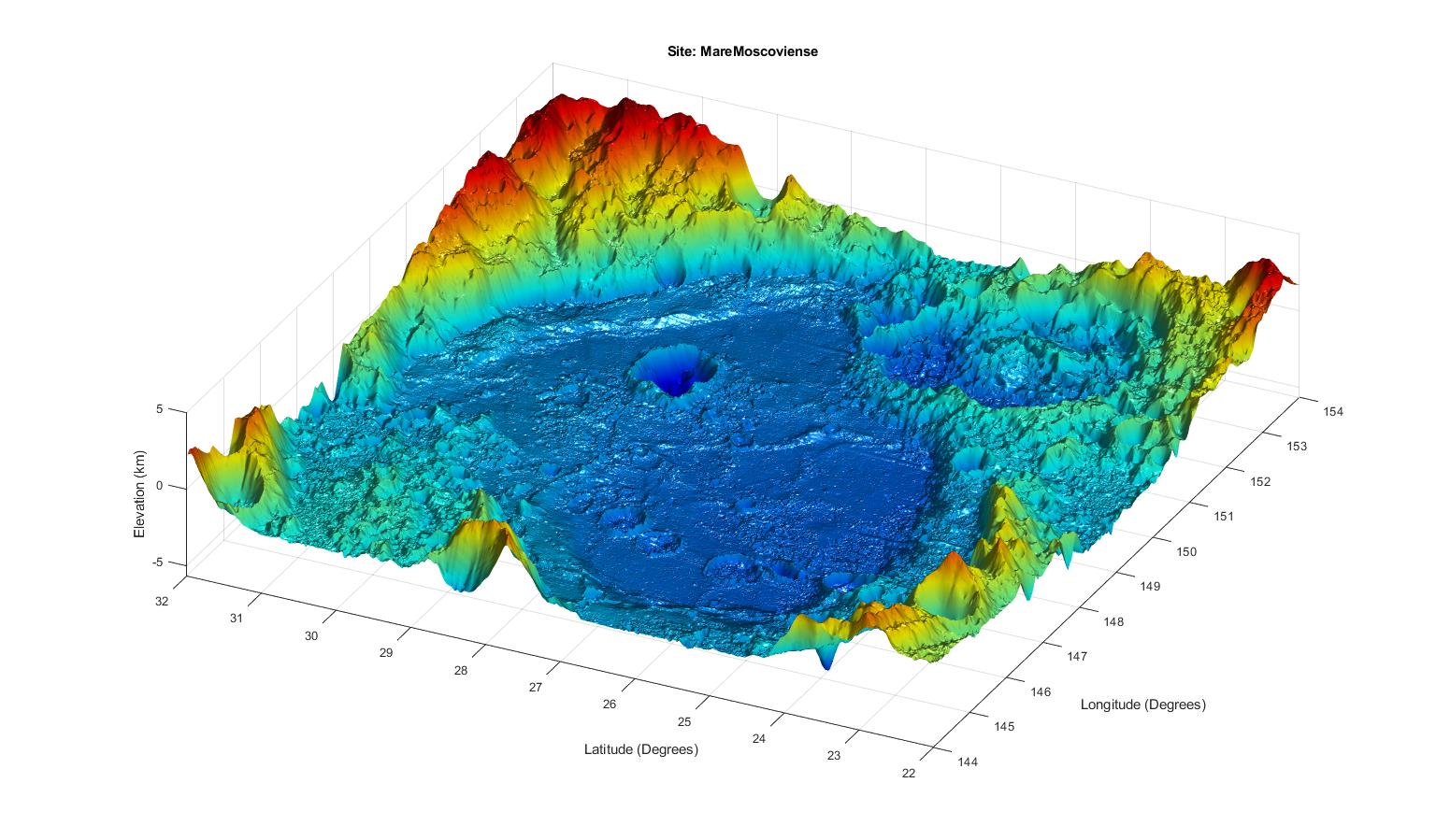}
        \caption{}
    \end{subfigure}
    \caption{DEMs of the lunar site Mare Moscoviense, a $487 x 445$ km region. Elevation in km is represented by a colour scale with 0 km at the lunar equatorial radius. (a) Single LOLA tracks colour-coded by height. White stripes on the image display areas lacking LOLA coverage. (b) 200 m scale interpolated DEM of the same region. (c) The perspective view of a smaller region (300 km on a side), with a factor of six exaggerated vertical axes, is shown using the interpolated 200 m scale DEM.}
    \label{fig:DEM}
\end{figure*}

Figure \ref{fig:DEM} (left) shows a non-interpolated DEM of Mare Moscoviense. The maximum separation between tracks is 1.8 km. Figure \ref{fig:DEM} (middle) shows the interpolated DEM of Mare Moscoviense produced in GMT. The right of Figure \ref{fig:DEM} (generated using MATLAB) shows the interpolated DEM in an exaggerated perspective view showing highlights and shading for a given illumination angle (-45, 30) applied to all of the MATLAB figures. The vertical range is 11 km.  

\subsection{Slope Maps}
Three visualisations of the slope maps were created for each site for the recognition of inaccessible terrain. All of the interesting information for rover deployment of the radio array is at low slopes, up to $\sim 25$°, with $< 15$° being especially important. 

\begin{figure*}
    \centering
    \begin{subfigure}[b]{0.3\textwidth}
        \centering
        \includegraphics[width=\textwidth]{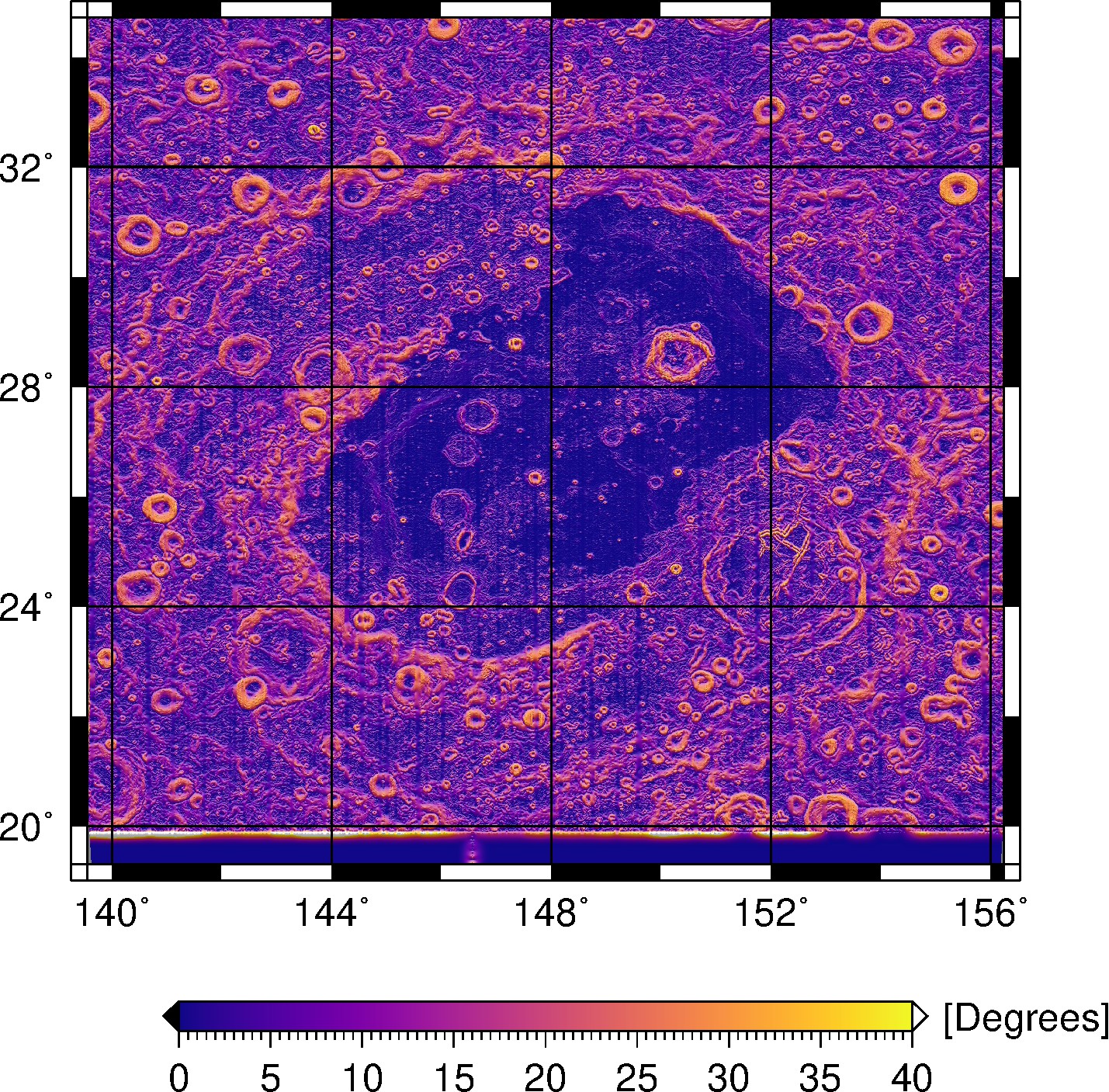}
        \caption{}
    \end{subfigure}
    \begin{subfigure}[b]{0.3\textwidth}
        \centering
        \includegraphics[width=\textwidth]{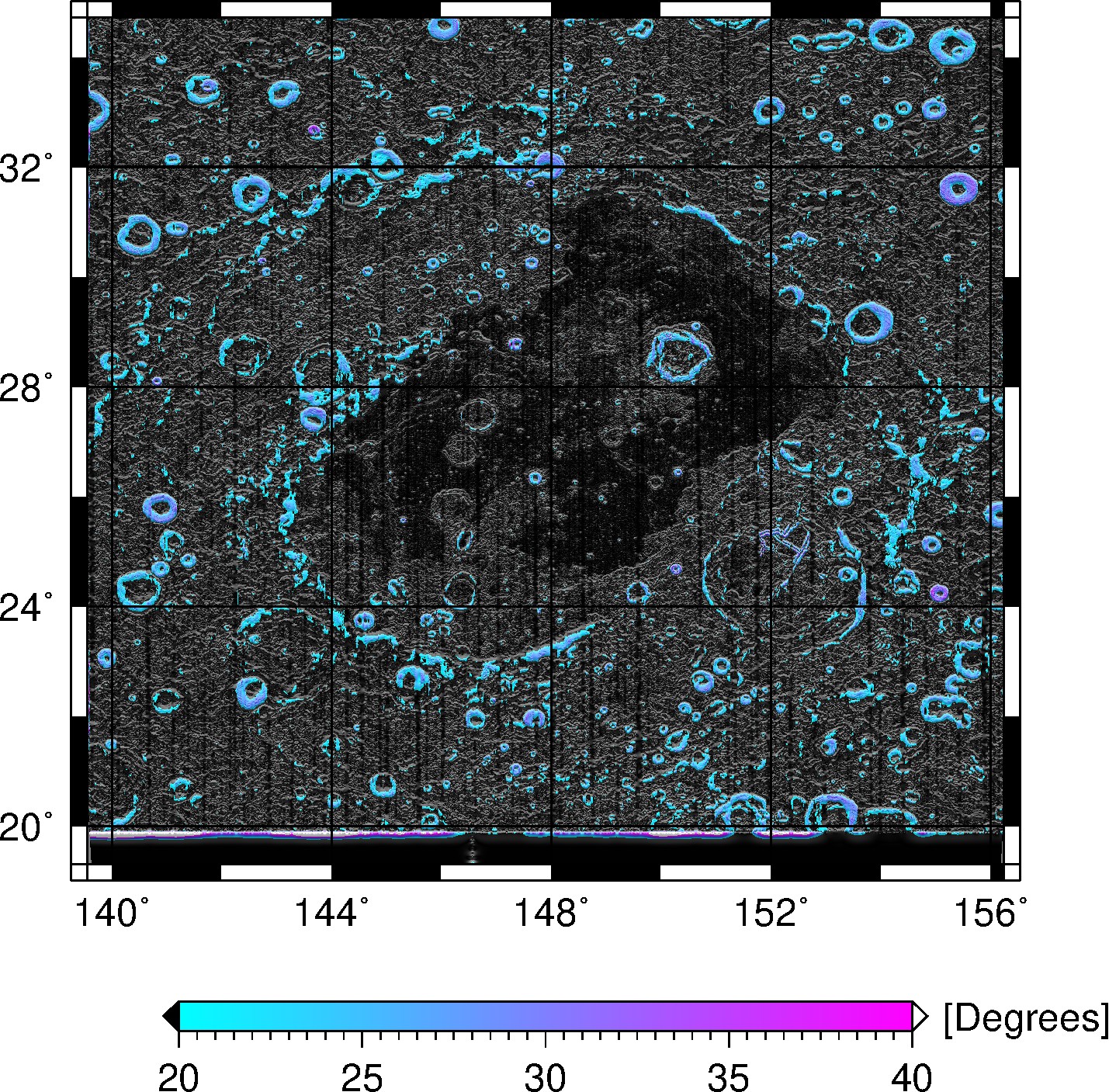}
        \caption{}
    \end{subfigure}
    \begin{subfigure}[b]{0.3\textwidth}
        \centering
        \includegraphics[width=\textwidth]{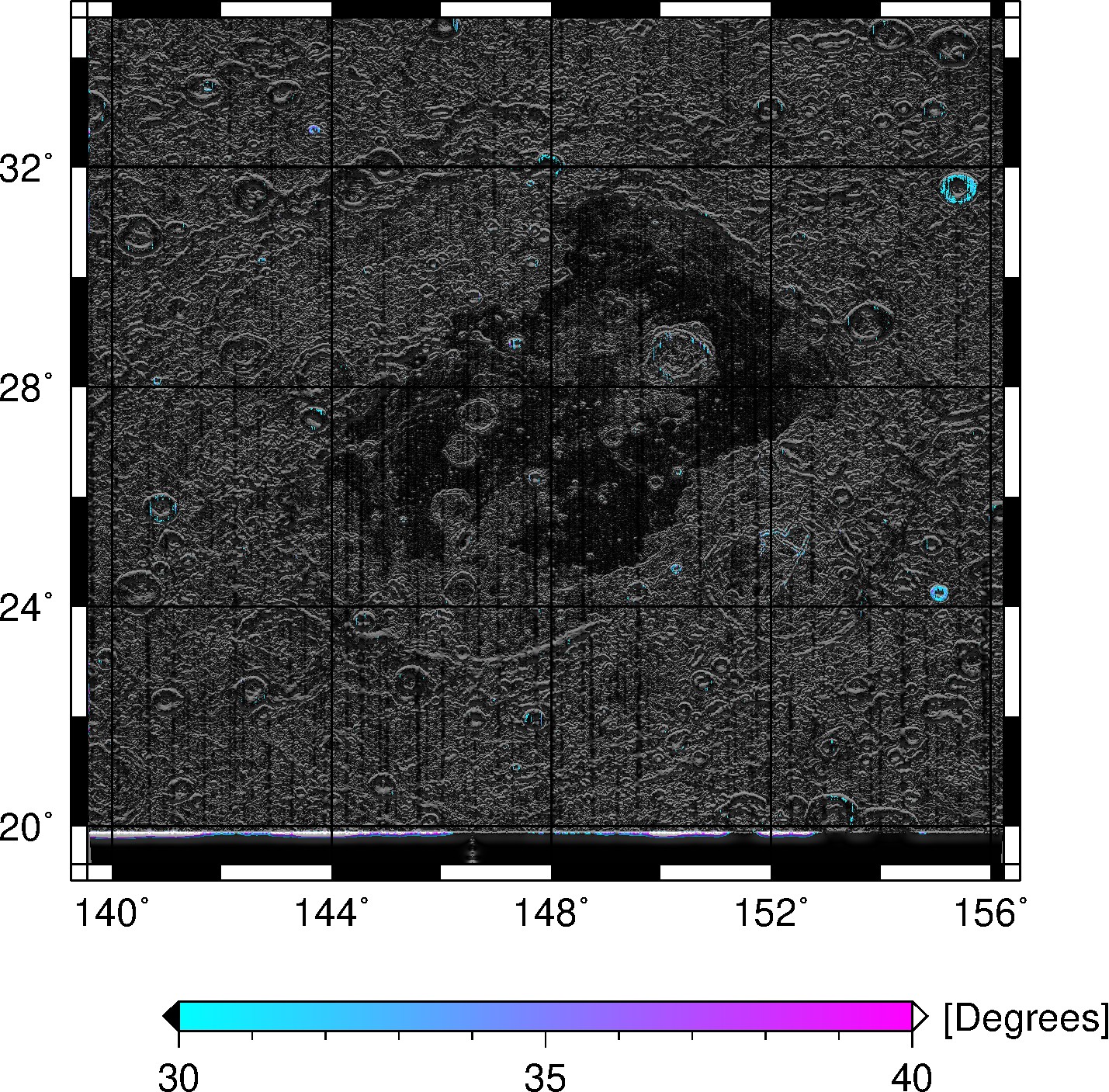}
        \caption{}
    \end{subfigure}
    \caption{Slope maps of the lunar site Mare Moscoviense, a $487 x 445$ km region. A colour scale represents the slope in degrees. (a) Slope over the range 0 - 40°. (b) Slope over the range of 20 - 40°. The slope below 20° is black. (c) Slope over the range of 30 - 40°. The slope below 30° is black.}
    \label{fig:D_Slope}
\end{figure*}

Figure \ref{fig:D_Slope} shows three slope maps of the site Mare Moscoviense with decreasing slope range scales. The mare floor does not have slopes greater than 20°. The middle map highlights the accessibility of the site, with all areas in black being less than 20° but identifies the smaller crater within the mare. The right map colours slope greater than 30°, inaccessible to all wheeled vehicles. Mare Moscoviense does not present challenges by a rather generous 30° criterion at the 200 m scale for wheeled vehicles. Even for a threshold of 15°, Mare Moscoviense is traversable except for the lower West region of the crater.

\subsection{Roughness Maps}
Maps on different scales were created for different topographic goals. 500 m scales measure hills and mountains and indicate possible flat terrain areas. 200 m and 100 m scales locate smaller site features, such as small impact craters but are subject to errors due to poor interpolation between tracks.

\begin{figure*}
    \centering
    \begin{subfigure}[b]{0.3\textwidth}
        \centering
        \includegraphics[width=\textwidth]{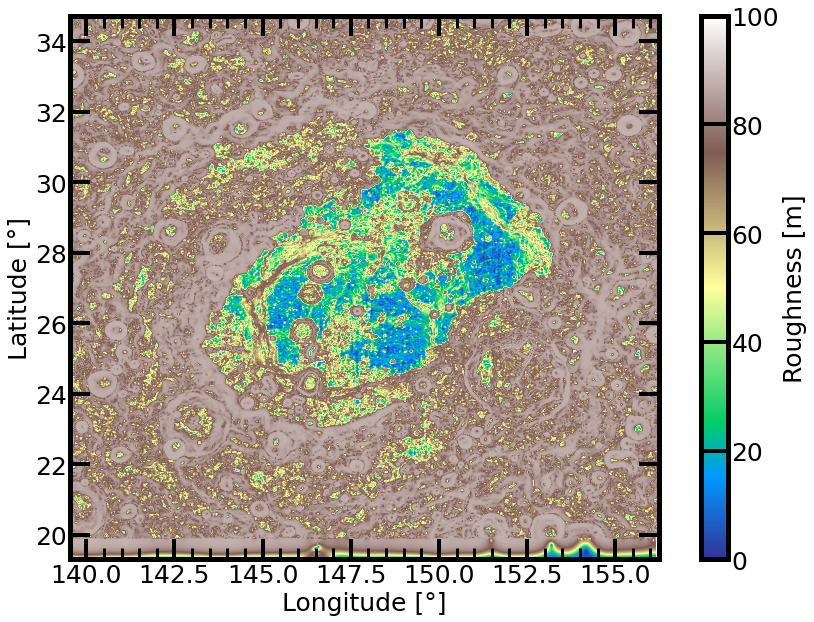}
        \caption{}
    \end{subfigure}
    \begin{subfigure}[b]{0.3\textwidth}
        \centering
        \includegraphics[width=\textwidth]{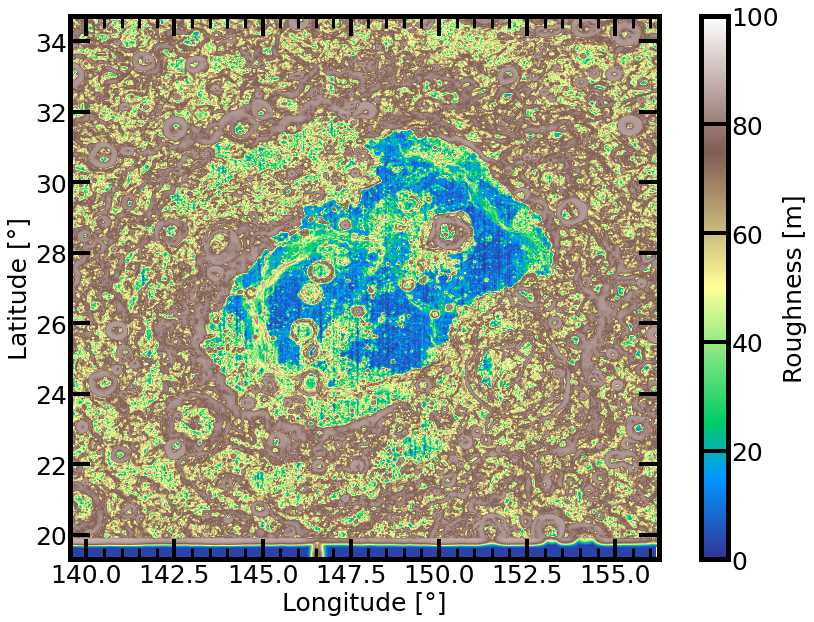}
        \caption{}
    \end{subfigure}
    \begin{subfigure}[b]{0.3\textwidth}
        \centering
        \includegraphics[width=\textwidth]{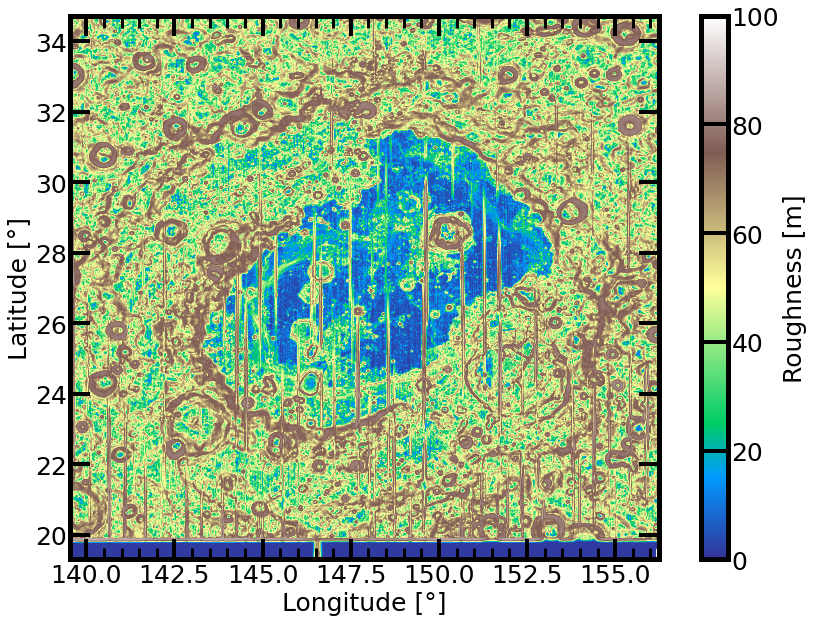}
        \caption{}
    \end{subfigure}
    \caption{RMS Roughness maps, ranging from 0 - 100 m, of the lunar site Mare Moscoviense on three scales, a $487 x 445$ km region, 500 m, 200 m, and 100 m scaled interpolated DEMs. Roughness in m is represented by a colour bar. (a) 500 m scale roughness map. (b) 200 m scale roughness map. (c) 100 m scale roughness map.}
    \label{fig:A_RMS}
\end{figure*}

Mare Moscoviense demonstrates the value of RMS roughness maps on these different scales. Figure \ref{fig:A_RMS} presents roughness maps of Mare Moscoviense on 500 m, 200 m and 100 m scales. A significant increase in low roughness (coloured green and blue in the maps) shows that smoother areas appear at smaller scales, as do erroneous tracks. The 500 m scale shows that Mare Moscoviense is a site to be studied in more detail because of a lower roughness area, though these still involved 50 m -scale roughness that would be impassable to a rover unless smaller scales show paths through. This terrain spans $\approx 280$ km, but smaller rough terrain within the area is present.

The 200 m scale shows an increase in less rough terrain (less than 20 m elevation changes) with isolated rough crater features.

The 100 m scale shows the prominent rough features at the mare walls. However, at this scale, track features due to interpolation are visible. The areas with low roughness (< 20 m roughness in 200 m scale are blue in Figure \ref{fig:A_RMS}) are more patchy than the slope < 20° areas. The low roughness area in the lower left region of the crater is divided by a linear feature of high roughness ($\sim$ 20 - 60 m). Whether this feature is traversable will require higher-resolution mapping.

The Gini coefficient can quantify the concentration of roughness levels. The Gini coefficient is dependent on the mean of the absolute difference between pairs of individual measures. The Gini coefficient ranges between zero and one for the clumping of roughness measurements. If the absolute difference between neighbouring measures of roughness is small (large), then the terrain is described as having constant (inconstant) roughness, and the Gini coefficient will be close to one (zero).

Topographic features such as these imply that rovers deploying radio antennae will have to take more circuitous paths to reach the full extent of the array. As a result, the rovers will have a larger payload and deployment time will increase.

\section{Site Comparison}
\label{sec:fom}
\subsection{Threshold Slope Maps}
Maps are tailored to characterise the suitability of the sites for a radio array produced from the maps described in Section \ref{sec:Results}. A threshold was applied to the slope maps to highlight inaccessible and highly accessible areas clarifying the differences between the sites for deploying an array.

The slope from roughness measurements is calculated by projecting the roughness value across the scale. A series of slope thresholds are applied to data sets which mask data exceeding the limit. The result is a slope map presenting the terrain likely traversable by a rover and indicating the accessibility of a site. 

The slope threshold boundaries were chosen by the maximum slope capability of successful rover missions listed in Table \ref{tab:rovers}. The maximum incline wheeled vehicles are designed for is 30° therefore, this is the maximum slope value in the scale. A conservative 25° is determined as the inaccessible slope limit because it allows several sites to qualify. A threshold of 15° is a safe choice, as all rovers can handle inclines up to this value. VIPER is planned to land on the Moon in late 2024 and has demonstrated complications traversing slopes > 15° (\citet{VIPER}). 15° is the lower limit to the threshold slope range.

\begin{table*}
\caption{Mars and Moon missions which included the deployment of rovers. The mission, rover name, location and landing date are listed. The rover properties: wheel diameter in cm; the maximum speed of the rover on flat terrain in $cms^{-1}$; the maximum allowed incline in degrees are listed.}
\begin{threeparttable}
\begin{tabular}{c|c|c|c|c|c|c}
\hline
Mission & Rover & Location & Landing Date & Wheel Diameter (cm) & Flat Speed ($cms^{-1}$) & Slope (degrees) \\
\hline
Luna 17                  & Lunokhod 1\footnote{}      & Moon     & 17/11/1970   & 26                  & 3                         &                         \\
American Apollo Program  & Lunar Roving Vehicle\footnote{}    & Moon     & 1971         & 41                  & 500                       & 15                      \\
Luna 21                  & Lunokhod 2\footnote{}         & Moon     & 16/01/1973   & 26                  & 28                        &                         \\
Chang'e 3                & Yutu\footnote{}          & Moon     & 12/14/2013   &                     &                           &                         \\
Chang'e 4                & Yutu-2\footnote{}        & Moon     & 01/03/2019   & 15                  & 6                         & 20                      \\
Mars Pathfinder        & Sojourner\footnote{}        & Mars     & 04/07/1997   & 6                   & 1                         &                         \\
Mars Surface Exploration & Spirit\footnote{}    & Mars     & 04/01/2004   & 13                  & 5                         & 30                      \\
Mars Surface Exploration & Opportunity\footnote{}          & Mars     & 24/01/2004   & 13                  & 5                         & 30                      \\
Mars Science Laboratory  & Curiosity\footnote{}        & Mars     & 05/08/2012    28                  & 4                         & 30                      \\
Mars 2020                & Perseverance\footnote{}   & Mars     & 18/02/2021   & 26                  & 2                         & 30\\ 
CLPS    &   VIPER\footnote{}    &   Moon    &   2024    &  50  &    20  &   15 \\

\hline
\end{tabular}
\begin{tablenotes}
    \small
    \item \textbf{3} http://selena.sai.msu.ru/Home/Spacecrafts/Lunokhod1/lunokhod1e.htm. \textbf{4} https://www.hq.nasa.gov/alsj/lrvhand.html. \textbf{5} http://www.orbitalfocus.uk/Diaries/Luna/Luna21.php. \textbf{6/7} https://spaceflight101.com/change/change-3/. \textbf{8} https://www.jpl.nasa.gov/missions/mars-pathfinder-sojourner-rover. \textbf{9} https://mars.nasa.gov/mer/. \textbf{10} https://mars.nasa.gov/mer/mission/rover-status/$\#opportunity$. \textbf{11} https://mars.nasa.gov/msl/home/. \textbf{12} https://mars.nasa.gov/mars2020/. \textbf{13} https://www.nasa.gov/viper.
\end{tablenotes}
\end{threeparttable}
\label{tab:rovers}
\end{table*}

\begin{figure*}
    \centering
    \includegraphics[width=\textwidth,height=0.9\textheight]{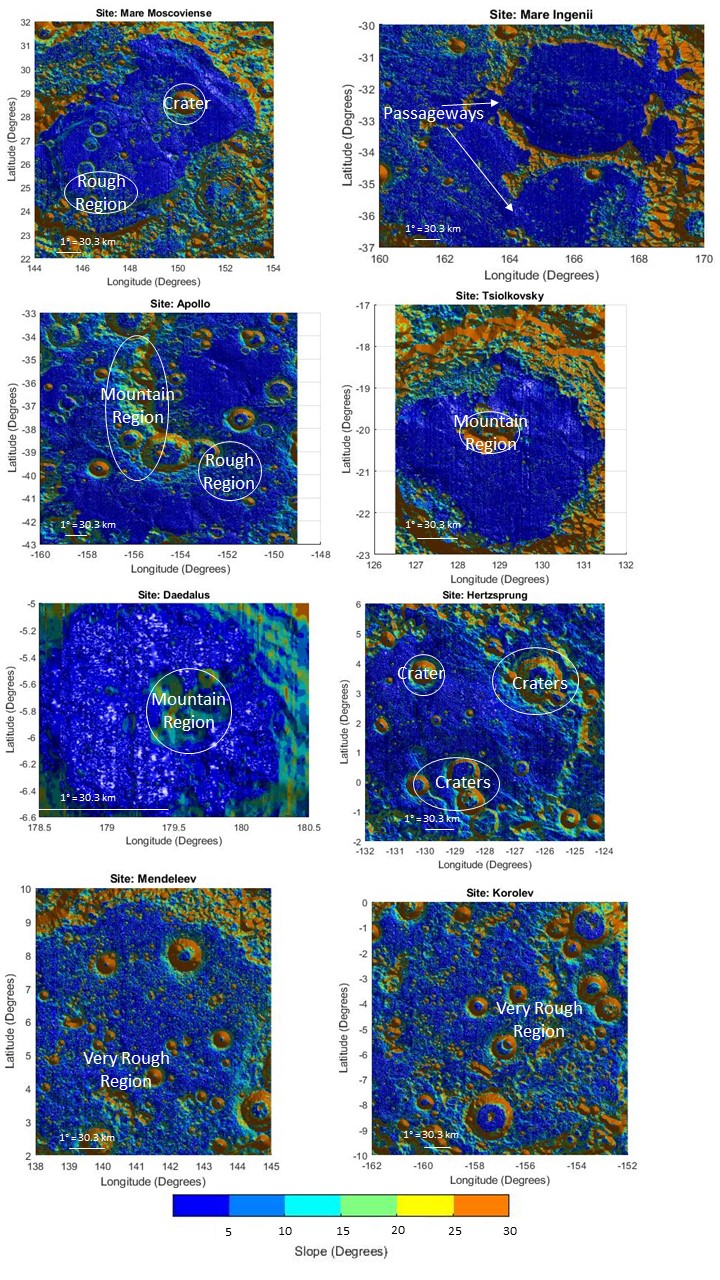}
    \caption{Site slope maps on the 200 m scale created from interpolated DEM. Slope in degrees is represented by a colour scale in increments of 5°. Blue areas are accessible. Green and yellow areas are challenging areas and orange is inaccessible to rovers. Special features are labelled.}
    \label{fig:all_sites}
\end{figure*}

Figure \ref{fig:all_sites} shows slope maps produced in Section \ref{sec:Slope} for all eight lunar sites. A colour scale represents the slope in increments of 5° rather than masked. Orange terrain represents inaccessibility (slope > 25°). Accessible and easily traversable terrain in blue. All the sites have large areas with slopes < 15°. Several of the sites have areas of rough terrain that will have to be avoided. Mare Ingenii is a striking example where the full extent of the site can only be accessed by passing through two narrow passes between ridges. Korolev appears less traversable, given the large areas of rough terrain. The varying difficulty of the terrain at each site is a factor in ranking the suitability of the sites.

\subsection{Figure of Merit (FoM)}
An objective comparison of these lunar sites can be helped by creating a suitable FoM that combines four factors: (i) Slope constrained size; (ii) Point Spread Function (PSF); (iii) roughness; (iv) terrain. These are discussed in turn below. 
\subsubsection{Factors}
Four factors were considered to assess the suitability of the sites of interest. 
\begin{enumerate}
    \item Slope-constrained site size. The size below a given slope threshold is the length of accessible terrain to a wheeled vehicle. A wheeled vehicle requires connected accessible terrain with routes around special features and the general roughness of a surface to deploy an array of dipoles. A larger site provides more flexibility in array design and a smaller PSF.
    \item PSF (angular resolution). Baseline lengths are the longest linear distances achievable on a site, representing the baselines of possible arrays. The length does not represent the true path length which considers avoiding objects and excessive terrain sloping, resulting in a greater distance traversed by the rover. At increasing slope thresholds, a site becomes more accessible and the achievable baseline increases. For the simplest array design, orthogonal baselines are required, but this is not attainable in some sites; bisecting baselines within 20° were used for this FoM.
    \item Roughness. If a surface is rough with rapidly varying elevation, then the Gini coefficient will be close to zero, and if the terrain is smooth with little elevation variation, the Gini measure is close to one. A highly ranked site will have a high Gini coefficient, indicating a more concentrated smooth surface.
    \item Terrain obstacles. Special features are identified by using interactive three-dimensional slope maps. Significant mountain regions, large obscuring craters, and the passageways around such features are special features. The questions asked are: Do these obstacles lie in the path of a dipole? Is extra distance traversed and extra cable (and greater mass) required to avoid these obstacles? 
\end{enumerate}

\subsubsection{Calculating Figures of Merit}
\begin{figure}
    \centering
    \includegraphics[width=\columnwidth]{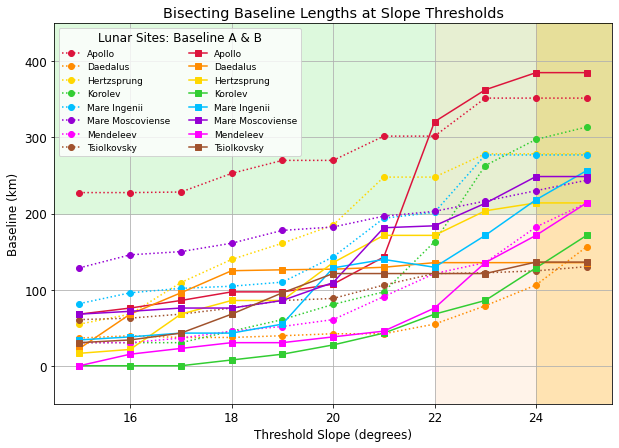}
    \caption{The maximum baseline (dotted line, A) and the maximum bisecting baseline (solid line, B) change for increasing slope thresholds, from 15° to 25°. Different coloured lines represent lunar sites. Lengths greater than 200 km are shaded in green. Slope threshold 22° - 24° is lightly shaded in orange, and slope thresholds greater than 24° are heavily shaded in orange.}
    \label{fig:thresh_slope}
\end{figure}

The FoM factors are listed in Table \ref{tab:FOM}. For a site at a given slope threshold, a point for each factor, between one (worst) and five (best), is awarded contingent on the value of each factor, as in Table \ref{tab:FOM}. Low slope and roughness on the meter scale are most important for a cosmology radio interferometer with a high concentration of dipoles in the central region, meaning the surface must be flat. Hence, the points for baseline (slope dependent) and roughness are doubled. A high overall score indicates a good site; the highest possible score is 30. Site size, baseline lengths and length ratios depend on the slope threshold, i.e. the awarded merit typically increases for increasing threshold slope. Gini is independent of the slope threshold, i.e. the awarded merit is the same value for all slope threshold maps. An example: Apollo, with an applied 20° slope threshold, has the longest length baseline $> 200$ km (compare Figure \ref{fig:thresh_slope}) and a bisecting baseline between the lengths 100 to 150 km, so the points awarded respectively are ten (factor awarded double points) and three from Table \ref{tab:FOM}.

\begin{table*}
    \caption{Factors contributing to the figure of merit and their awarded points for each slope threshold. Factors with double points are annotated with $\ast$.}
    \label{tab:FOM}
    \begin{tabular}{c|c|c|c|c}
    \hline
      & i. & ii. & ii. & iii. %& iv.v. & vi.
    \\
    Points Awarded & Site Size (km) & Baseline (km)$\ast$ & Baseline Ratio & Roughness (Gini)$\ast$
    \\
    \hline
    5 & \textgreater 300 & \textgreater 200 & 1:1  & > 0.4 
    \\
    4 & 250 - 300 & 150 - 200 & 1:2 & 0.3 - 0.4 
    \\
    3 & 200 - 250 & 100 - 150 & 1:3 & 0.2 - 0.3 
    \\
    2 & 150 - 200 & 50 - 100 & 1:4 & 0.1 - 0.2 
    \\
    1 & \textless 150 & 0 - 50 & 1:5 & 0.0 - 0.1 
    \\
    \hline
    \end{tabular}
\end{table*}
The figure of merit is calculated for a given site by summing the points awarded for each factor for a given slope threshold. Two baseline points are awarded for the two baselines which bisect. The maximum number of points awarded by the figure of merit is 30. Below, each factor is analysed and followed by the figure of merit result in Table \ref{tab:FOM_results}.

Figure \ref{fig:thresh_slope} shows how the maximum baseline (dotted line) and the maximum bisecting baseline change (dashed line) for increasing slope thresholds from 15° to 25°. For some sites, the longest achievable baselines depend strongly on threshold slope, e.g., Apollo's second baseline. Sites are shown in different colours. The 22° slope threshold for a bisecting baseline has only one site, Apollo, with a length greater than 200 km, but by increasing the slope threshold to 24°, four sites have lengths with the required 200 km length. The area between 22° and 24° is shaded in light orange to highlight the changes in baselines, and baselines $> 200$ km are shaded in green.

The simplest radio interferometer design would involve two bisecting orthogonal dipoles, each 200 km in diameter. For such a design, it is important for the dipole lengths to be close to a 1:1 ratio. The site Korolev does not achieve a 1:1 ratio, whereas all other sites are close to a 1:1 ratio with the most tolerant slope limit, 25°.

Roughness and slope maps of Korolev suggest that the site is one of the roughest and least desirable sites for an interferometer by eye. The Gini coefficient validates this assumption by being the minimum, 0.23. Similarly, sites showing the most promise from observing their topographic maps are Apollo, Mare Ingenii, and Daedalus. These sites have the highest Gini coefficients $> 0.3$, implying more consistent terrains.

\begin{figure*}
    \centering
    \includegraphics[width=\textwidth]{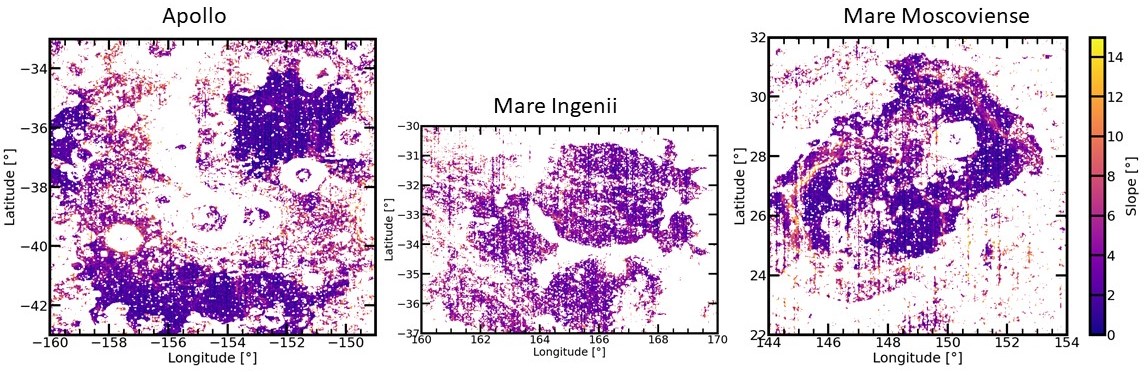}
    \caption{Left to the right, slope maps of the three highest ranked lunar sites Apollo ($338 x 306$ km), Mare Ingenii ($306 x 213 km$) and Mare Moscoviense ($306 x 306$ km). A colour scale with a 15° slope threshold represents the slope in degrees. White areas cannot be traversed and purple are the smoothest surfaces.}
    \label{fig:15deg}
\end{figure*}

With the FoM characteristics considered, the weighting of each characteristic is summed, and the total merit for each site is calculated. The highest-ranked site is determined by having the greatest total points. Tables \ref{tab:fom24} and \ref{tab:FOM_results} show the figure of merit results by presenting the measured factor value and the sum of points awarded at each slope threshold, respectively. The most feasible locations for a 200 km array are Apollo, Mare Ingenii and Mare Moscoviense. The threshold slopes 15°, 20°, 22° and 24° were chosen. The lower boundary, 15°, is not sufficient to traverse a 200 km baseline, except in Apollo. The following thresholds show increases in the number of sites achieving a 200 km baseline. Figure \ref{fig:15deg} shows the 15° threshold map of the top-ranked sites Apollo, Mare Ingenii and Mare Moscoviense. A restrictive limit allows for one baseline to be traversed. However, a near perpendicular bisecting baseline cannot be traversed.
\begin{table*}
    \centering
    \begin{tabular}{c|c|c|c|c|c|c|c|c|c|c|c|c|c}
    \hline
    Sites & \multicolumn{4}{c}{Site Size (km)} & \multicolumn{4}{c}{Baseline (km)} & \multicolumn{4}{c}{Baseline Ratio} & Roughness (Gini) 
    \\
      & 15° & 20° & 22° & 24° & 15° & 20° & 22° & 24° & 15° & 20° & 22° & 24° &  \\
    \hline
    Apollo & 228 & 270 & 320 & 385 & 67 & 107 & 301 & 352 & 3.36 & 2.52 & 0.94 & 0.91 & 0.40
    \\
    Mare Ingenii & 82 & 143 & 202 & 277 & 34 & 129 & 129 & 218 & 2.41 & 1.11 & 1.56 & 1.27 & 0.37
    \\
    Mare Moscoviense & 129 & 182 & 203 & 249 & 68 & 109 & 184 & 230 & 1.90 & 1.67 & 1.10 & 0.93 & 0.34
    \\
    Hertzsprung & 55 & 185 & 248 & 278 & 16 & 135 & 171 & 214 & 3.35 & 1.36 & 1.45 & 1.30 & 0.32
    \\
    Tsiolkovsky & 61 & 135 & 135 & 136 & 30 & 88 & 121 & 125 & 2.00 & 0.73 & 1.00 & 0.96 & 0.36
    \\
    Daedalus & 36 & 121 & 135 & 135 & 23 & 42 & 55 & 106 & 1.60 & 0.33 & 0.40 & 0.78 & 0.40
    \\
    Mendeleev & 30 & 60 & 121 & 182 & 0 & 38 & 76 & 171 & - & 1.60 & 1.60 & 1.06 & 0.27
    \\
    Korolev & 30 & 81 & 162 & 297 & 0 & 27 & 68 & 129 & - & 2.95 & 2.40 & 2.31 & 0.23
    \\
    \hline
    \end{tabular}
    \caption{Figure of merit for all eight sites at a 15°, 20°, 22° and 24° slope thresholds. The factor values in this table determine the merit points awarded (points awarded between one to five shown in Table \ref{tab:FOM}). Table \ref{tab:FOM_results} gives the sum of points awarded for each of the eight sites at each slope threshold.} \label{tab:fom24}
\end{table*}

\begin{table} 
    \centering
    \caption{Figure of merit results. The figure of merit is calculated for each site at the slope thresholds 15°, 20°, 22° and 24°. The maximum figure of merit value is 30. The table is ordered by the highest-ranked site at 20° first to the lowest-ranked site last. \label{tab:FOM_results}}
    \begin{tabular}{c|c|c|c|c}
    \hline
    Site & \multicolumn{4}{c}{Threshold Slope}\\
     & 15° & 20° & 22° & 24° \\
    \hline
    Apollo & 25 & 26 & 30 & 30 \\
    Mare Ingenii & 19 & 24 & 27 & 30 \\
    Mare Moscoviense & 20 & 23 & 27 & 28 \\
    Hertzsprung & 14 & 22 & 25 & 26 \\
    Tsiolkovsky & 19 & 22 & 24 & 24 \\
    Daedalus & 17 & 19 & 21 & 24 \\
    Mendeleev & 9 & 11 & 14 & 19 \\
    Korolev & 6 & 10 & 15 & 19 \\
    \hline
    \end{tabular}
\end{table}

\section{Discussion}
\label{sec:Discussion}
We conducted a preliminary lunar far-side site survey to determine the optimum location for a 200 km radio-class, low-frequency (10s MHz) interferometer. The lunar far-side has few smooth craters and mare regions where a 200 km radio interferometer could be situated. LOLA Digital terrain maps and slope and roughness maps derived from them were used to examine eight potential sites. The surface of the Moon is a strenuous environment for wheeled vehicles. The engineering limits of the existing and proposed vehicles influence the threshold slope of terrain maps. Since we measured an average slope on a 200 m scale, slope thresholds of 15°, 22° and 24° were used. 22° and 24° show an increase in sites with 200 km baselines. It should be noted that a 15° slope threshold is not sufficient to achieve the goal of a 200 km baseline interferometer.

We examined four factors: slope-constrained size, PSF, roughness and terrain obstacles. From the DTMs, slope maps and roughness maps, measurements of these factors created the FoM to describe and rank the sites. Terrain features affect the ease of deployment of the antennas of the array. Figure \ref{fig:all_sites} shows the significant features of each site and their accessibility.

Apollo is ranked top in the FoM because it is the largest site, having the largest baseline through terrain with minimal slope. However, the largest baseline is challenging because of the mountainous terrain bisecting the crater. The roughness Gini coefficient of Apollo is high because the crater floor is a smooth U-shaped region. Notably, within the U-shape of accessible terrain, a highly rough region in the lower right region of the crater poses a challenge to a rover (see also Figure \ref{fig:15deg}). 

Mare Moscoviense has few obscuring objects within the mare floor but does show rough regions, though these appear avoidable. 

Mare Ingenii has an entirely flat, $< 5$°, surface separated by a steep mountain wall $> 25$°. Between crater walls are flat passageways, which create accessible routes for a rover. 

If the interferometer maximum baseline could be reduced whilst achieving the scientific goal, Tsiolkovsky and Daedalus become suitable locations for arrays smaller than 150 km. The two sites have the most accessible terrain within their crater floors and show one avoidable mountainous feature greater than the threshold slope. 

The sites Hertzsprung, Korolev and Mendeleev show very rough and disrupted terrain with features inclined greater than 25°. Routes through rougher sites are not identifiable and challenging because the 200 m scale on which the maps were interpolated implies that non-mapped features would be present. 

Limitations to this study arose from both computational resources and data anomalies. The topographic maps shown in this paper were produced with the best achievable scale of 200 m per pixel. The LOLA instrument has a nominal ranging precision of 10 cm and a vertical precision of < 1 m (\cite{Smith2017}). LOLA can calculate slope on the smallest scale $\sim 25$ m. The anomaly affects the instrument's operation by the common case of two spots out of five emitted being detected. To overcome this challenge, interpolated data was created to estimate the surface between track coverage and the greater scale was chosen for computational efficiency. Using large scales means we do not have a detailed understanding of the lunar surface at these sites, and unobserved obstacles could make a site inaccessible. \cite{FERNANDES2021} generated $\sim$ 1 m per pixel scale topographic maps of the site Mare Ingenii using images from the LRO Camera, which capture shadows cast from sunlight, and they relate this to the gradient. Future work includes using high-resolution topographic maps to determine realistic traverses and resultant cable length requirements. Improvements to the FoM can be made. For example, different weights can be assigned to each factor depending on how they affect mission design requirements (e.g., traverse length, available baseline given realistic slope capabilities).

\subsection{Site Protection}
This study has identified the importance of a few lunar far-side sites with the likelihood that these locations would be highly competitive in the next decades. The lack of smooth terrain on the far-side increases the value of such exceptional sites. In this case, preventing harmful interference at the sites will form disputes over entitlement to access sites regardless of the local resources (\citet{Elvis2021}). Currently, profitable lunar sites correspond to ‘common-pool resources’ (\citet{Edwards1998}) in which ‘no single nation has a generally recognized exclusive jurisdiction’ (\citet{Wijkman1982}). The impact of the SpaceX Starlink Satellites has been observed on the Zwicky Transient Facility Survey observations (\citet{Prezemek2022}). Growing concerns over the impact on astronomical observations from low Earth orbit satellite constellations demonstrate the urgency to protect future research from similar experiences (\citet{Lawrence2021}). We expect to see developments within the field as a recently established group within the International Academy of Astronautics (IAA), the Moon Farside Protection Permanent Committee, aims to call attention to lunar interference corruption\footnote{See https://iaaspace.org/about/permanent-committees/\#1658152007849-7e3d453e-d9ba}. 

\section{Conclusions}
\label{sec:Conclusion}
We have conducted a preliminary study of lunar far-side radio array sites.
\begin{enumerate}
    \item Eight sites of dimension > 100 km were investigated by generating topographic maps on the scale of 200 m with a ranging precision of 10 cm and a vertical precision of < 1 m;
    \item Only the site Apollo is traversable (meaning a linear 200 km baseline can be accessed by a rover) when a 15° slope threshold is applied to the region;
    \item Four sites (adding Mare Ingenii, Mare Moscoviense, and Hertzsprung) are traversable for a 24° slope threshold;
    \item The sites Tsiolkovsky and Daedalus would be good sites for smaller arrays $\sim$ 100 km;
    \item A figure of merit was created using size, slope, roughness, and topographic features to compare the sites objectively.
\end{enumerate}
The ongoing work to achieve higher-resolution topography maps would provide a more rigorous site study. The rarity of good sites points to a need for their protection.

\section*{Acknowledgements}
This work was supported by STFC [ST/X508354/1]. This research used GMT,\footnote{https://www.generic-mapping-tools.org/} a free, open-source software licensed under the LGPL \citep{Wessel2019}. This research also made use of MATLAB,\footnote{https://www.mathworks.com/products/matlab.html?s\_tid=hp\_products\_matlab} a programming and numeric computing platform \citep{MATLAB:2021b}. This research also used Astropy,\footnote{http://www.astropy.org} a community-developed core Python package for Astronomy \citep{Astropy2013, Astropy2018}. I would like to thank Diego Altamirano and Jeremy Drake for their support of this project.

\section*{Data Availability}
The data in this research can be found in the PDS Geosciences Node Lunar Orbital Data Explorer (ODE)\footnote{https://ode.rsl.wustl.edu/moon/index.aspx}.

\bibliographystyle{rasti}
\bibliography{references}

\begin{thebibliography}{63}
\expandafter\ifx\csname natexlab\endcsname\relax\def\natexlab#1{#1}\fi

\bibitem[Adams et~al.(1993)Adams, Bond, Freese, Frieman, \& Olinto]{Adams1993}
Adams, F.~C., Bond, J.~R., Freese, K., Frieman, J.~A., \& Olinto, A.~V., 1993.
\newblock Natural inflation: Particle physics models, power-law spectra for
  large-scale structure, and constraints from the cosmic background explorer,
  {\it Phys. Rev. D\/}, {\bf 47}, 426--455.

\bibitem[Araki et~al.(2009)Araki, Tazawa, Noda, Ishihara, Goossens, Sasaki,
  Kawano, Kamiya, Otake, Oberst, \& Shum]{Araki2009}
Araki, H., Tazawa, S., Noda, H., Ishihara, Y., Goossens, S., Sasaki, S.,
  Kawano, N., Kamiya, I., Otake, H., Oberst, J., \& Shum, C., 2009.
\newblock Lunar global shape and polar topography derived from kaguya-lalt
  laser altimetry, {\it Science\/}, {\bf 323}(5916), 897--900.

\bibitem[{Astropy Collaboration} et~al.(2013){Astropy Collaboration},
  {Robitaille}, {Tollerud}, {Greenfield}, {Droettboom}, {Bray}, {Aldcroft},
  {Davis}, {Ginsburg}, {Price-Whelan}, {Kerzendorf}, {Conley}, {Crighton},
  {Barbary}, {Muna}, {Ferguson}, {Grollier}, {Parikh}, {Nair}, {Unther},
  {Deil}, {Woillez}, {Conseil}, {Kramer}, {Turner}, {Singer}, {Fox}, {Weaver},
  {Zabalza}, {Edwards}, {Azalee Bostroem}, {Burke}, {Casey}, {Crawford},
  {Dencheva}, {Ely}, {Jenness}, {Labrie}, {Lim}, {Pierfederici}, {Pontzen},
  {Ptak}, {Refsdal}, {Servillat}, \& {Streicher}]{Astropy2013}
{Astropy Collaboration}, {Robitaille}, T.~P., {Tollerud}, E.~J., {Greenfield},
  P., {Droettboom}, M., {Bray}, E., {Aldcroft}, T., {Davis}, M., {Ginsburg},
  A., {Price-Whelan}, A.~M., {Kerzendorf}, W.~E., {Conley}, A., {Crighton}, N.,
  {Barbary}, K., {Muna}, D., {Ferguson}, H., {Grollier}, F., {Parikh}, M.~M.,
  {Nair}, P.~H., {Unther}, H.~M., {Deil}, C., {Woillez}, J., {Conseil}, S.,
  {Kramer}, R., {Turner}, J. E.~H., {Singer}, L., {Fox}, R., {Weaver}, B.~A.,
  {Zabalza}, V., {Edwards}, Z.~I., {Azalee Bostroem}, K., {Burke}, D.~J.,
  {Casey}, A.~R., {Crawford}, S.~M., {Dencheva}, N., {Ely}, J., {Jenness}, T.,
  {Labrie}, K., {Lim}, P.~L., {Pierfederici}, F., {Pontzen}, A., {Ptak}, A.,
  {Refsdal}, B., {Servillat}, M., \& {Streicher}, O., 2013.
\newblock {Astropy: A community Python package for astronomy}, {\it \aap\/},
  {\bf 558}, A33.

\bibitem[{Astropy Collaboration} et~al.(2018){Astropy Collaboration},
  {Price-Whelan}, {Sip{\H{o}}cz}, {G{\"u}nther}, {Lim}, {Crawford}, {Conseil},
  {Shupe}, {Craig}, {Dencheva}, {Ginsburg}, {VanderPlas}, {Bradley},
  {P{\'e}rez-Su{\'a}rez}, {de Val-Borro}, {Aldcroft}, {Cruz}, {Robitaille},
  {Tollerud}, {Ardelean}, {Babej}, {Bach}, {Bachetti}, {Bakanov}, {Bamford},
  {Barentsen}, {Barmby}, {Baumbach}, {Berry}, {Biscani}, {Boquien}, {Bostroem},
  {Bouma}, {Brammer}, {Bray}, {Breytenbach}, {Buddelmeijer}, {Burke},
  {Calderone}, {Cano Rodr{\'\i}guez}, {Cara}, {Cardoso}, {Cheedella}, {Copin},
  {Corrales}, {Crichton}, {D'Avella}, {Deil}, {Depagne}, {Dietrich}, {Donath},
  {Droettboom}, {Earl}, {Erben}, {Fabbro}, {Ferreira}, {Finethy}, {Fox},
  {Garrison}, {Gibbons}, {Goldstein}, {Gommers}, {Greco}, {Greenfield},
  {Groener}, {Grollier}, {Hagen}, {Hirst}, {Homeier}, {Horton}, {Hosseinzadeh},
  {Hu}, {Hunkeler}, {Ivezi{\'c}}, {Jain}, {Jenness}, {Kanarek}, {Kendrew},
  {Kern}, {Kerzendorf}, {Khvalko}, {King}, {Kirkby}, {Kulkarni}, {Kumar},
  {Lee}, {Lenz}, {Littlefair}, {Ma}, {Macleod}, {Mastropietro}, {McCully},
  {Montagnac}, {Morris}, {Mueller}, {Mumford}, {Muna}, {Murphy}, {Nelson},
  {Nguyen}, {Ninan}, {N{\"o}the}, {Ogaz}, {Oh}, {Parejko}, {Parley}, {Pascual},
  {Patil}, {Patil}, {Plunkett}, {Prochaska}, {Rastogi}, {Reddy Janga},
  {Sabater}, {Sakurikar}, {Seifert}, {Sherbert}, {Sherwood-Taylor}, {Shih},
  {Sick}, {Silbiger}, {Singanamalla}, {Singer}, {Sladen}, {Sooley},
  {Sornarajah}, {Streicher}, {Teuben}, {Thomas}, {Tremblay}, {Turner},
  {Terr{\'o}n}, {van Kerkwijk}, {de la Vega}, {Watkins}, {Weaver}, {Whitmore},
  {Woillez}, {Zabalza}, \& {Astropy Contributors}]{Astropy2018}
{Astropy Collaboration}, {Price-Whelan}, A.~M., {Sip{\H{o}}cz}, B.~M.,
  {G{\"u}nther}, H.~M., {Lim}, P.~L., {Crawford}, S.~M., {Conseil}, S.,
  {Shupe}, D.~L., {Craig}, M.~W., {Dencheva}, N., {Ginsburg}, A., {VanderPlas},
  J.~T., {Bradley}, L.~D., {P{\'e}rez-Su{\'a}rez}, D., {de Val-Borro}, M.,
  {Aldcroft}, T.~L., {Cruz}, K.~L., {Robitaille}, T.~P., {Tollerud}, E.~J.,
  {Ardelean}, C., {Babej}, T., {Bach}, Y.~P., {Bachetti}, M., {Bakanov}, A.~V.,
  {Bamford}, S.~P., {Barentsen}, G., {Barmby}, P., {Baumbach}, A., {Berry},
  K.~L., {Biscani}, F., {Boquien}, M., {Bostroem}, K.~A., {Bouma}, L.~G.,
  {Brammer}, G.~B., {Bray}, E.~M., {Breytenbach}, H., {Buddelmeijer}, H.,
  {Burke}, D.~J., {Calderone}, G., {Cano Rodr{\'\i}guez}, J.~L., {Cara}, M.,
  {Cardoso}, J.~V.~M., {Cheedella}, S., {Copin}, Y., {Corrales}, L.,
  {Crichton}, D., {D'Avella}, D., {Deil}, C., {Depagne}, {\'E}., {Dietrich},
  J.~P., {Donath}, A., {Droettboom}, M., {Earl}, N., {Erben}, T., {Fabbro}, S.,
  {Ferreira}, L.~A., {Finethy}, T., {Fox}, R.~T., {Garrison}, L.~H., {Gibbons},
  S.~L.~J., {Goldstein}, D.~A., {Gommers}, R., {Greco}, J.~P., {Greenfield},
  P., {Groener}, A.~M., {Grollier}, F., {Hagen}, A., {Hirst}, P., {Homeier},
  D., {Horton}, A.~J., {Hosseinzadeh}, G., {Hu}, L., {Hunkeler}, J.~S.,
  {Ivezi{\'c}}, {\v{Z}}., {Jain}, A., {Jenness}, T., {Kanarek}, G., {Kendrew},
  S., {Kern}, N.~S., {Kerzendorf}, W.~E., {Khvalko}, A., {King}, J., {Kirkby},
  D., {Kulkarni}, A.~M., {Kumar}, A., {Lee}, A., {Lenz}, D., {Littlefair},
  S.~P., {Ma}, Z., {Macleod}, D.~M., {Mastropietro}, M., {McCully}, C.,
  {Montagnac}, S., {Morris}, B.~M., {Mueller}, M., {Mumford}, S.~J., {Muna},
  D., {Murphy}, N.~A., {Nelson}, S., {Nguyen}, G.~H., {Ninan}, J.~P.,
  {N{\"o}the}, M., {Ogaz}, S., {Oh}, S., {Parejko}, J.~K., {Parley}, N.,
  {Pascual}, S., {Patil}, R., {Patil}, A.~A., {Plunkett}, A.~L., {Prochaska},
  J.~X., {Rastogi}, T., {Reddy Janga}, V., {Sabater}, J., {Sakurikar}, P.,
  {Seifert}, M., {Sherbert}, L.~E., {Sherwood-Taylor}, H., {Shih}, A.~Y.,
  {Sick}, J., {Silbiger}, M.~T., {Singanamalla}, S., {Singer}, L.~P., {Sladen},
  P.~H., {Sooley}, K.~A., {Sornarajah}, S., {Streicher}, O., {Teuben}, P.,
  {Thomas}, S.~W., {Tremblay}, G.~R., {Turner}, J.~E.~H., {Terr{\'o}n}, V.,
  {van Kerkwijk}, M.~H., {de la Vega}, A., {Watkins}, L.~L., {Weaver}, B.~A.,
  {Whitmore}, J.~B., {Woillez}, J., {Zabalza}, V., \& {Astropy Contributors},
  2018.
\newblock {The Astropy Project: Building an Open-science Project and Status of
  the v2.0 Core Package}, {\it \aj\/}, {\bf 156}(3), 123.

\bibitem[{Barker} et~al.(2021){Barker}, {Mazarico}, {Neumann}, {Smith},
  {Zuber}, \& {Head}]{Barker2021}
{Barker}, M.~K., {Mazarico}, E., {Neumann}, G.~A., {Smith}, D.~E., {Zuber},
  M.~T., \& {Head}, J.~W., 2021.
\newblock {Improved LOLA elevation maps for south pole landing sites: Error
  estimates and their impact on illumination conditions}, {\it \planss\/}, {\bf
  203}, 105119.

\bibitem[{Bassett} et~al.(2020){Bassett}, {Rapetti}, {Burns}, {Tauscher}, \&
  {MacDowall}]{Bassett2020}
{Bassett}, N., {Rapetti}, D., {Burns}, J.~O., {Tauscher}, K., \& {MacDowall},
  R., 2020.
\newblock {Characterizing the radio quiet region behind the lunar farside for
  low radio frequency experiments}, {\it Advances in Space Research\/}, {\bf
  66}(6), 1265--1275.

\bibitem[Bowman et~al.(2018)Bowman, Rogers, Monsalve, Mozdzen, \&
  Mahesh]{Bowman2018}
Bowman, J.~D., Rogers, A. E.~E., Monsalve, R.~A., Mozdzen, T.~J., \& Mahesh,
  N., 2018.
\newblock An absorption profile centred at 78 megahertz in the sky-averaged
  spectrum, {\it Nature\/}, {\bf 555}(7694), 67--70.

\bibitem[Burns(2021)]{Burns_21}
Burns, J., 2021.
\newblock Transformative science from the lunar farside: observations of the
  dark ages and exoplanetary systems at low radio frequencies, {\it
  Philosophical Transactions of the Royal Society A: Mathematical, Physical and
  Engineering Sciences\/}, {\bf 379}, 20190564.

\bibitem[Burns et~al.(2021)Burns, MacDowall, Bale, Hallinan, Bassett, \&
  Hegedus]{Burns2021}
Burns, J.~O., MacDowall, R., Bale, S., Hallinan, G., Bassett, N., \& Hegedus,
  A., 2021.
\newblock Low radio frequency observations from the moon enabled by {NASA}
  landed payload missions, {\it The Planetary Science Journal\/}, {\bf 2}(2),
  44.

\bibitem[Bussey et~al.(2003)Bussey, Lucey, Steutel, Robinson, Spudis, \&
  Edwards]{Bussey2003}
Bussey, D. B.~J., Lucey, P.~G., Steutel, D., Robinson, M.~S., Spudis, P.~D., \&
  Edwards, K.~D., 2003.
\newblock Permanent shadow in simple craters near the lunar poles, {\it
  Geophysical Research Letters\/}, {\bf 30}(6).

\bibitem[Chin et~al.(2007)Chin, Brylow, Foote, Garvin, Kasper, Keller, Litvak,
  Mitrofanov, Paige, Raney, Robinson, Sanin, Smith, Spence, Spudis, Stern, \&
  Zuber]{Chin2007}
Chin, G., Brylow, S., Foote, M., Garvin, J., Kasper, J., Keller, J., Litvak,
  M., Mitrofanov, I., Paige, D., Raney, K., Robinson, M., Sanin, A., Smith, D.,
  Spence, H., Spudis, P., Stern, S., \& Zuber, M., 2007.
\newblock Lunar reconnaissance orbiter overview: The instrument suite and
  mission, {\it Space Science Reviews\/}, {\bf 129}, 391--419.

\bibitem[Council(1997)]{NRC1997}
Council, N.~R., 1997.
\newblock {\it Lessons Learned from the Clementine Mission\/}, The National
  Academies Press, Washington, DC.

\bibitem[{Dabhade} et~al.(2020){Dabhade}, {R{\"o}ttgering}, {Bagchi},
  {Shimwell}, {Hardcastle}, {Sankhyayan}, {Morganti}, {Jamrozy}, {Shulevski},
  \& {Duncan}]{Dabhade2020}
{Dabhade}, P., {R{\"o}ttgering}, H.~J.~A., {Bagchi}, J., {Shimwell}, T.~W.,
  {Hardcastle}, M.~J., {Sankhyayan}, S., {Morganti}, R., {Jamrozy}, M.,
  {Shulevski}, A., \& {Duncan}, K.~J., 2020.
\newblock {Giant radio galaxies in the LOFAR Two-metre Sky Survey. I. Radio and
  environmental properties}, {\it \aap\/}, {\bf 635}, A5.

\bibitem[{de Gasperin} et~al.(2021){de Gasperin}, {Williams, W. L.}, {Best,
  P.}, {Br\"uggen, M.}, {Brunetti, G.}, {Cuciti, V.}, {Dijkema, T. J.},
  {Hardcastle, M. J.}, {Norden, M. J.}, {Offringa, A.}, {Shimwell, T.}, {van
  Weeren, R.}, {Bomans, D.}, {Bonafede, A.}, {Botteon, A.}, {Callingham, J.
  R.}, {Cassano, R.}, {Chyzy, K. T.}, {Emig, K. L.}, {Edler, H.}, {Haverkorn,
  M.}, {Heald, G.}, {Heesen, V.}, {Iacobelli, M.}, {Intema, H. T.}, {Kadler,
  M.}, {Malek, K.}, {Mevius, M.}, {Miley, G.}, {Mingo, B.}, {Morabito, L. K.},
  {Sabater, J.}, {Morganti, R.}, {Orr\'u, E.}, {Pizzo, R.}, {Prandoni, I.},
  {Shulevski, A.}, {Tasse, C.}, {Vaccari, M.}, {Zarka, P.}, \& {R\"ottgering,
  H.}]{Gasperin2021}
{de Gasperin}, {Williams, W. L.}, {Best, P.}, {Br\"uggen, M.}, {Brunetti, G.},
  {Cuciti, V.}, {Dijkema, T. J.}, {Hardcastle, M. J.}, {Norden, M. J.},
  {Offringa, A.}, {Shimwell, T.}, {van Weeren, R.}, {Bomans, D.}, {Bonafede,
  A.}, {Botteon, A.}, {Callingham, J. R.}, {Cassano, R.}, {Chyzy, K. T.},
  {Emig, K. L.}, {Edler, H.}, {Haverkorn, M.}, {Heald, G.}, {Heesen, V.},
  {Iacobelli, M.}, {Intema, H. T.}, {Kadler, M.}, {Malek, K.}, {Mevius, M.},
  {Miley, G.}, {Mingo, B.}, {Morabito, L. K.}, {Sabater, J.}, {Morganti, R.},
  {Orr\'u, E.}, {Pizzo, R.}, {Prandoni, I.}, {Shulevski, A.}, {Tasse, C.},
  {Vaccari, M.}, {Zarka, P.}, \& {R\"ottgering, H.}, 2021.
\newblock The lofar lba sky survey - i. survey description and preliminary data
  release, {\it A\&A\/}, {\bf 648}, A104.

\bibitem[Edwards \& Steins(1998)]{Edwards1998}
Edwards, V.~M. \& Steins, N.~A., 1998.
\newblock Developing an analytical framework for multiple-use commons, {\it
  Journal of Theoretical Politics\/}, {\bf 10}, 347 -- 383.

\bibitem[Elvis et~al.(2021)Elvis, Krolikowski, \& Milligan]{Elvis2021}
Elvis, M., Krolikowski, A., \& Milligan, T., 2021.
\newblock Concentrated lunar resources: imminent implications for governance
  and justice, {\it Philosophical Transactions of the Royal Society A:
  Mathematical, Physical and Engineering Sciences\/}, {\bf 379}(2188),
  20190563.

\bibitem[{Fanaroff} \& {Riley}(1974)]{Fanaroff1974}
{Fanaroff}, B.~L. \& {Riley}, J.~M., 1974.
\newblock {The morphology of extragalactic radio sources of high and low
  luminosity}, {\it \mnras\/}, {\bf 167}, 31P--36P.

\bibitem[{Fanti} et~al.(1990){Fanti}, {Fanti}, {Schilizzi}, {Spencer}, {Nan
  Rendong}, {Parma}, {van Breugel}, \& {Venturi}]{Fanti1990}
{Fanti}, R., {Fanti}, C., {Schilizzi}, R.~T., {Spencer}, R.~E., {Nan Rendong},
  {Parma}, P., {van Breugel}, W.~J.~M., \& {Venturi}, T., 1990.
\newblock {On the nature of compact steep spectrum radio sources.}, {\it
  \aap\/}, {\bf 231}, 333--346.

\bibitem[Fernandes \& Mosegaard(2022)]{FERNANDES2021}
Fernandes, I. \& Mosegaard, K., 2022.
\newblock High-resolution topography from planetary images and laser altimetry,
  {\it Planetary and Space Science\/}, {\bf 218}, 105514.

\bibitem[Furlanetto et~al.(2006)Furlanetto, {Peng Oh}, \&
  Briggs]{Furlanetto2006}
Furlanetto, S.~R., {Peng Oh}, S., \& Briggs, F.~H., 2006.
\newblock Cosmology at low frequencies: The 21cm transition and the
  high-redshift universe, {\it Physics Reports\/}, {\bf 433}(4), 181--301.

\bibitem[G.A.(2009)]{Neumann2009}
G.A., N., 2009.
\newblock Lunar orbiter laser altimeter raw data set,
  \url{https://pds.nasa.gov/ds-view/pds/viewProfile.jsp?dsid=LRO-L-LOLA-4-GDR-V1.0}.

\bibitem[Gläser(2014)]{Glaser2014}
Gläser, P., 2014.
\newblock {\it Evaluation of topography, slopes, illumination and surface
  roughness of landing sites near the lunar south pole using laser altimetry
  from the lunar reconnaissance orbiter\/}, Ph.D. thesis, Technischen
  Universit¨at Berlin.

\bibitem[Gläser et~al.(2013)Gläser, Haase, Oberst, \& Neumann]{Glaser2013}
Gläser, P., Haase, I., Oberst, J., \& Neumann, G., 2013.
\newblock Co-registration of laser altimeter tracks with digital terrain models
  and applications in planetary science, {\it Planetary and Space Science\/},
  {\bf 89}.

\bibitem[Gläser et~al.(2017)Gläser, Oberst, Neumann, Mazarico, Speyerer, \&
  Robinson]{Glaser2017}
Gläser, P., Oberst, J., Neumann, G., Mazarico, E., Speyerer, E., \& Robinson,
  M., 2017.
\newblock Illumination conditions at the lunar poles: Implications for future
  exploration, {\it Planetary and Space Science\/}, {\bf 162}.

\bibitem[Kaiser \& Weiler(2000)]{Kaiser2000}
Kaiser, M.~L. \& Weiler, K.~W., 2000.
\newblock {\it The Current Status of Low Frequency Radio Astronomy from
  Space\/}, pp. 1--11, American Geophysical Union (AGU).

\bibitem[Kim(2021)]{Kim2021}
Kim, J., 2021.
\newblock Modeling the tidal locking mechanism of the moon in function of
  inverse-square law and hooke's law, in {\em 11th IEEE Annual Computing and
  Communication Workshop and Conference, CCWC 2021, Las Vegas, NV, USA, January
  27-30, 2021\/}, pp. 1070--1073, IEEE.

\bibitem[{Koopmans} et~al.(2021){Koopmans}, {Barkana}, {Bentum}, {Bernardi},
  {Boonstra}, {Bowman}, {Burns}, {Chen}, {Datta}, {Falcke}, {Fialkov},
  {Gehlot}, {Gurvits}, {Jeli{\'c}}, {Klein-Wolt}, {Lazio}, {Meerburg},
  {Mellema}, {Mertens}, {Mesinger}, {Offringa}, {Pritchard}, {Semelin},
  {Subrahmanyan}, {Silk}, {Trott}, {Vedantham}, {Verde}, {Zaroubi}, \&
  {Zarka}]{2021Koopmans}
{Koopmans}, L. V.~E., {Barkana}, R., {Bentum}, M., {Bernardi}, G., {Boonstra},
  A.-J., {Bowman}, J., {Burns}, J., {Chen}, X., {Datta}, A., {Falcke}, H.,
  {Fialkov}, A., {Gehlot}, B., {Gurvits}, L., {Jeli{\'c}}, V., {Klein-Wolt},
  M., {Lazio}, J., {Meerburg}, D., {Mellema}, G., {Mertens}, F., {Mesinger},
  A., {Offringa}, A., {Pritchard}, J., {Semelin}, B., {Subrahmanyan}, R.,
  {Silk}, J., {Trott}, C., {Vedantham}, H., {Verde}, L., {Zaroubi}, S., \&
  {Zarka}, P., 2021.
\newblock {Peering into the dark (ages) with low-frequency space
  interferometers}, {\it Experimental Astronomy\/}, {\bf 51}(3), 1641--1676.

\bibitem[Kouloumvakos et~al.(2021)Kouloumvakos, Rouillard, Warmuth, Magdalenic,
  Jebaraj, Mann, Vainio, \& Monstein]{Kouloumvakos_2021}
Kouloumvakos, A., Rouillard, A., Warmuth, A., Magdalenic, J., Jebaraj, I.~C.,
  Mann, G., Vainio, R., \& Monstein, C., 2021.
\newblock Coronal conditions for the occurrence of type ii radio bursts, {\it
  The Astrophysical Journal\/}, {\bf 913}(2), 99.

\bibitem[Kreslavsky \& Head~III(2000)]{Kreslavsky2000}
Kreslavsky, M.~A. \& Head~III, J.~W., 2000.
\newblock Kilometer-scale roughness of mars: Results from mola data analysis,
  {\it Journal of Geophysical Research: Planets\/}, {\bf 105}(E11),
  26695--26711.

\bibitem[Lawrence(2021)]{Lawrence2021}
Lawrence, A., 2021.
\newblock {\it Losing the Sky\/}, Photon Productions.

\bibitem[Loeb \& Zaldarriaga(2004)]{Loeb2004}
Loeb, A. \& Zaldarriaga, M., 2004.
\newblock Measuring the small-scale power spectrum of cosmic density
  fluctuations through 21 cm tomography prior to the epoch of structure
  formation, {\it Physical review letters\/}, {\bf 92}, 211301.

\bibitem[Maccone(2021)]{Maccone2021}
Maccone, C., 2021.
\newblock Moon farside protection, moon village and pac (protected antipode
  circle), in {\em The Search for ExtraTerrestrial Intelligence\/}, pp. 75--84,
  Springer International Publishing, Cham.

\bibitem[{Mao} et~al.(2008){Mao}, {Tegmark}, {McQuinn}, {Zaldarriaga}, \&
  {Zahn}]{Mao2008}
{Mao}, Y., {Tegmark}, M., {McQuinn}, M., {Zaldarriaga}, M., \& {Zahn}, O.,
  2008.
\newblock {How accurately can 21cm tomography constrain cosmology?}, {\it
  \prd\/}, {\bf 78}(2), 023529.

\bibitem[MATLAB(2021)]{MATLAB:2021b}
MATLAB, 2021.
\newblock {\it R2021b\/}, The MathWorks Inc., Natick, Massachusetts.

\bibitem[McKean et~al.(2016)McKean, Godfrey, Vegetti, Wise, Morganti,
  Hardcastle, Rafferty, Anderson, Avruch, Beck, Bell, van Bemmel, Bentum,
  Bernardi, Best, Blaauw, Bonafede, Breitling, Broderick, Brüggen, Cerrigone,
  Ciardi, de~Gasperin, Deller, Duscha, Engels, Falcke, Fallows, Frieswijk,
  Garrett, Grießmeier, van Haarlem, Heald, Hoeft, Horst, Iacobelli, Intema,
  Juette, Karastergiou, Kondratiev, Koopmans, Kuniyoshi, Kuper, van Leeuwen,
  Maat, Mann, Markoff, McFadden, McKay-Bukowski, Mulcahy, Munk, Nelles, Orru,
  Paas, Pandey-Pommier, Pietka, Pizzo, Polatidis, Reich, Röttgering,
  Rowlinson, Scaife, Serylak, Shulevski, Sluman, Smirnov, Steinmetz, Stewart,
  Swinbank, Tagger, Thoudam, Toribio, Vermeulen, Vocks, van Weeren, Wucknitz,
  Yatawatta, \& Zarka]{McKean2016}
McKean, J.~P., Godfrey, L. E.~H., Vegetti, S., Wise, M.~W., Morganti, R.,
  Hardcastle, M.~J., Rafferty, D., Anderson, J., Avruch, I.~M., Beck, R., Bell,
  M.~E., van Bemmel, I., Bentum, M.~J., Bernardi, G., Best, P., Blaauw, R.,
  Bonafede, A., Breitling, F., Broderick, J.~W., Brüggen, M., Cerrigone, L.,
  Ciardi, B., de~Gasperin, F., Deller, A., Duscha, S., Engels, D., Falcke, H.,
  Fallows, R.~A., Frieswijk, W., Garrett, M.~A., Grießmeier, J.~M., van
  Haarlem, M.~P., Heald, G., Hoeft, M., Horst, A. J. v.~d., Iacobelli, M.,
  Intema, H., Juette, E., Karastergiou, A., Kondratiev, V.~I., Koopmans, L.
  V.~E., Kuniyoshi, M., Kuper, G., van Leeuwen, J., Maat, P., Mann, G.,
  Markoff, S., McFadden, R., McKay-Bukowski, D., Mulcahy, D.~D., Munk, H.,
  Nelles, A., Orru, E., Paas, H., Pandey-Pommier, M., Pietka, M., Pizzo, R.,
  Polatidis, A.~G., Reich, W., Röttgering, H. J.~A., Rowlinson, A., Scaife, A.
  M.~M., Serylak, M., Shulevski, A., Sluman, J., Smirnov, O., Steinmetz, M.,
  Stewart, A., Swinbank, J., Tagger, M., Thoudam, S., Toribio, M.~C.,
  Vermeulen, R., Vocks, C., van Weeren, R.~J., Wucknitz, O., Yatawatta, S., \&
  Zarka, P., 2016.
\newblock {LOFAR imaging of Cygnus A – direct detection of a turnover in the
  hotspot radio spectra}, {\it Monthly Notices of the Royal Astronomical
  Society\/}, {\bf 463}(3), 3143--3150.

\bibitem[Mebane et~al.(2020)Mebane, Mirocha, \& Furlanetto]{Mebane2020}
Mebane, R.~H., Mirocha, J., \& Furlanetto, S.~R., 2020.
\newblock The effects of population {III} radiation backgrounds on the
  cosmological 21-cm signal, {\it Monthly Notices of the Royal Astronomical
  Society\/}, {\bf 493}(1), 1217--1226.

\bibitem[{Mr{\'o}z} et~al.(2022){Mr{\'o}z}, {Otarola}, {Prince}, {Dekany},
  {Duev}, {Graham}, {Groom}, {Masci}, \& {Medford}]{Prezemek2022}
{Mr{\'o}z}, P., {Otarola}, A., {Prince}, T.~A., {Dekany}, R., {Duev}, D.~A.,
  {Graham}, M.~J., {Groom}, S.~L., {Masci}, F.~J., \& {Medford}, M.~S., 2022.
\newblock {Impact of the SpaceX Starlink Satellites on the Zwicky Transient
  Facility Survey Observations}, {\it \apjl\/}, {\bf 924}(2), L30.

\bibitem[{Mu{\~n}oz} et~al.(2015){Mu{\~n}oz}, {Ali-Ha{\"\i}moud}, \&
  {Kamionkowski}]{Munoz2015}
{Mu{\~n}oz}, J.~B., {Ali-Ha{\"\i}moud}, Y., \& {Kamionkowski}, M., 2015.
\newblock {Primordial non-gaussianity from the bispectrum of 21-cm fluctuations
  in the dark ages}, {\it \prd\/}, {\bf 92}(8), 083508.

\bibitem[{Muller} \& {Oort}(1951)]{Muller1951}
{Muller}, C.~A. \& {Oort}, J.~H., 1951.
\newblock {Observation of a Line in the Galactic Radio Spectrum: The
  Interstellar Hydrogen Line at 1,420 Mc./sec., and an Estimate of Galactic
  Rotation}, {\it \nat\/}, {\bf 168}(4270), 357--358.

\bibitem[NASA/GSFC(2010)]{NASA}
NASA/GSFC, 2010.
\newblock The lunar far side as seen by the lunar orbiter laser altimeter,
  \url{https://www.nasa.gov/mission_pages/LRO/multimedia/lroimages/lola-20100312-farside.html}.

\bibitem[{National Academies of Sciences, Engineering, and
  Medicine}(2021)]{NAP26141}
{National Academies of Sciences, Engineering, and Medicine}, 2021.
\newblock {\it {Pathways to Discovery in Astronomy and Astrophysics for the
  2020s}\/}, The National Academies Press, Washington, DC.

\bibitem[{Pilia} et~al.(2016){Pilia}, {Hessels}, {Stappers}, {Kondratiev},
  {Kramer}, {van Leeuwen}, {Weltevrede}, {Lyne}, {Zagkouris}, {Hassall},
  {Bilous}, {Breton}, {Falcke}, {Grie{\ss}meier}, {Keane}, {Karastergiou},
  {Kuniyoshi}, {Noutsos}, {Os{\l}owski}, {Serylak}, {Sobey}, {ter Veen},
  {Alexov}, {Anderson}, {Asgekar}, {Avruch}, {Bell}, {Bentum}, {Bernardi},
  {B{\^\i}rzan}, {Bonafede}, {Breitling}, {Broderick}, {Br{\"u}ggen}, {Ciardi},
  {Corbel}, {de Geus}, {de Jong}, {Deller}, {Duscha}, {Eisl{\"o}ffel},
  {Fallows}, {Fender}, {Ferrari}, {Frieswijk}, {Garrett}, {Gunst}, {Hamaker},
  {Heald}, {Horneffer}, {Jonker}, {Juette}, {Kuper}, {Maat}, {Mann}, {Markoff},
  {McFadden}, {McKay-Bukowski}, {Miller-Jones}, {Nelles}, {Paas},
  {Pandey-Pommier}, {Pietka}, {Pizzo}, {Polatidis}, {Reich}, {R{\"o}ttgering},
  {Rowlinson}, {Schwarz}, {Smirnov}, {Steinmetz}, {Stewart}, {Swinbank},
  {Tagger}, {Tang}, {Tasse}, {Thoudam}, {Toribio}, {van der Horst},
  {Vermeulen}, {Vocks}, {van Weeren}, {Wijers}, {Wijnands}, {Wijnholds},
  {Wucknitz}, \& {Zarka}]{Pilia2016}
{Pilia}, M., {Hessels}, J.~W.~T., {Stappers}, B.~W., {Kondratiev}, V.~I.,
  {Kramer}, M., {van Leeuwen}, J., {Weltevrede}, P., {Lyne}, A.~G.,
  {Zagkouris}, K., {Hassall}, T.~E., {Bilous}, A.~V., {Breton}, R.~P.,
  {Falcke}, H., {Grie{\ss}meier}, J.~M., {Keane}, E., {Karastergiou}, A.,
  {Kuniyoshi}, M., {Noutsos}, A., {Os{\l}owski}, S., {Serylak}, M., {Sobey},
  C., {ter Veen}, S., {Alexov}, A., {Anderson}, J., {Asgekar}, A., {Avruch},
  I.~M., {Bell}, M.~E., {Bentum}, M.~J., {Bernardi}, G., {B{\^\i}rzan}, L.,
  {Bonafede}, A., {Breitling}, F., {Broderick}, J.~W., {Br{\"u}ggen}, M.,
  {Ciardi}, B., {Corbel}, S., {de Geus}, E., {de Jong}, A., {Deller}, A.,
  {Duscha}, S., {Eisl{\"o}ffel}, J., {Fallows}, R.~A., {Fender}, R., {Ferrari},
  C., {Frieswijk}, W., {Garrett}, M.~A., {Gunst}, A.~W., {Hamaker}, J.~P.,
  {Heald}, G., {Horneffer}, A., {Jonker}, P., {Juette}, E., {Kuper}, G.,
  {Maat}, P., {Mann}, G., {Markoff}, S., {McFadden}, R., {McKay-Bukowski}, D.,
  {Miller-Jones}, J.~C.~A., {Nelles}, A., {Paas}, H., {Pandey-Pommier}, M.,
  {Pietka}, M., {Pizzo}, R., {Polatidis}, A.~G., {Reich}, W., {R{\"o}ttgering},
  H., {Rowlinson}, A., {Schwarz}, D., {Smirnov}, O., {Steinmetz}, M.,
  {Stewart}, A., {Swinbank}, J.~D., {Tagger}, M., {Tang}, Y., {Tasse}, C.,
  {Thoudam}, S., {Toribio}, M.~C., {van der Horst}, A.~J., {Vermeulen}, R.,
  {Vocks}, C., {van Weeren}, R.~J., {Wijers}, R.~A.~M.~J., {Wijnands}, R.,
  {Wijnholds}, S.~J., {Wucknitz}, O., \& {Zarka}, P., 2016.
\newblock {Wide-band, low-frequency pulse profiles of 100 radio pulsars with
  LOFAR}, {\it \aap\/}, {\bf 586}, A92.

\bibitem[Ratkowski \& Foster(2014)]{Ratkowski2014}
Ratkowski, R. \& Foster, J., 2014.
\newblock Libration of the moon.

\bibitem[Riris \& Cavanaugh(2009)]{Riris2009}
Riris, H. \& Cavanaugh, J., 2009.
\newblock Lunar reconnaissance orbiter project: Calibration document for the
  lunar orbiter laser altimeter (lola) instrument, {\it LUNAR RECONNAISSANCE
  ORBITER PROJECT\/}.

\bibitem[Robinson et~al.(2010)Robinson, Brylow, \& Tschimmel]{Robinson2010}
Robinson, M., Brylow, S., \& Tschimmel, M. e.~a., 2010.
\newblock Lunar reconnaissance orbiter camera (lroc) instrument overview., {\it
  Space Science Review\/}, {\bf 150}, 81–124.

\bibitem[Rosenburg et~al.(2011)Rosenburg, Aharonson, Head, Kreslavsky,
  Mazarico, Neumann, Smith, Torrence, \& Zuber]{Rosenburg2011}
Rosenburg, M.~A., Aharonson, O., Head, J.~W., Kreslavsky, M.~A., Mazarico, E.,
  Neumann, G.~A., Smith, D.~E., Torrence, M.~H., \& Zuber, M.~T., 2011.
\newblock Global surface slopes and roughness of the moon from the lunar
  orbiter laser altimeter, {\it Journal of Geophysical Research: Planets\/},
  {\bf 116}(E2).

\bibitem[Scholten et~al.(2011)Scholten, Oberst, Matz, Roatsch, Wählisch,
  Gläser, Robinson, Mazarico, Neumann, Zuber, \& Smith]{Scholten2011}
Scholten, F., Oberst, J., Matz, K.-D., Roatsch, T., Wählisch, M., Gläser, P.,
  Robinson, M., Mazarico, E., Neumann, G., Zuber, M., \& Smith, D., 2011.
\newblock in {\em Complementary LRO Global Lunar Topography Datasets —A
  Comparison of 100 Meter Raster DTMs from LROC WAC Stereo (GLD100) and LOLA
  Altimetry Data\/}, p. 2080.

\bibitem[Scholten et~al.(2012)Scholten, Oberst, Matz, Roatsch, Wählisch,
  Speyerer, \& Robinson]{Scholten2012}
Scholten, F., Oberst, J., Matz, K.-D., Roatsch, T., Wählisch, M., Speyerer,
  E.~J., \& Robinson, M.~S., 2012.
\newblock {\it Journal of Geophysical Research: Planets\/}, {\bf 117}(E12).

\bibitem[Shepard et~al.(2001)Shepard, Campbell, Bulmer, Farr, Gaddis, \&
  Plaut]{Shepard2001}
Shepard, M.~K., Campbell, B.~A., Bulmer, M.~H., Farr, T.~G., Gaddis, L.~R., \&
  Plaut, J.~J., 2001.
\newblock The roughness of natural terrain: A planetary and remote sensing
  perspective, {\it Journal of Geophysical Research: Planets\/}, {\bf
  106}(E12), 32777--32795.

\bibitem[{Shirley} et~al.(2022){Shirley}, {Balaban}, {Colaprete}, {Elphic},
  {Sanchez}, {Falcone}, {Beyer}, {Banerjee}, \& {Bradner}]{VIPER}
{Shirley}, M., {Balaban}, E., {Colaprete}, A., {Elphic}, R.~C., {Sanchez}, H.,
  {Falcone}, L., {Beyer}, R., {Banerjee}, S., \& {Bradner}, K., 2022.
\newblock {VIPER Traverse Planning}, in {\em 53rd Lunar and Planetary Science
  Conference\/}, vol. 2678 of {\bf LPI Contributions}, p. 2874.

\bibitem[{Shultz} \& {Spudis}(1983)]{Shultz1983}
{Shultz}, P.~H. \& {Spudis}, P.~D., 1983.
\newblock {Beginning and end of lunar mare volcanism}, {\it \nat\/}, {\bf
  302}(5905), 233--236.

\bibitem[Silk(1967)]{Silk1967}
Silk, J., 1967.
\newblock Fluctuations in the primordial fireball., {\it Nature (London), 215:
  1155-6(Sept. 9, 1967).\/}.

\bibitem[{Silk}(2018)]{Silk2018}
{Silk}, J., 2018.
\newblock {Towards the Limits of Cosmology}, {\it Foundations of Physics\/},
  {\bf 48}(10), 1305--1332.

\bibitem[{Singh} et~al.(2022){Singh}, {Nambissan}~T., {Subrahmanyan}, {Udaya
  Shankar}, {Girish}, {Raghunathan}, {Somashekar}, {Srivani}, \&
  {Sathyanarayana Rao}]{singh2022}
{Singh}, S., {Nambissan}~T., J., {Subrahmanyan}, R., {Udaya Shankar}, N.,
  {Girish}, B.~S., {Raghunathan}, A., {Somashekar}, R., {Srivani}, K.~S., \&
  {Sathyanarayana Rao}, M., 2022.
\newblock On the detection of a cosmic dawn signal in the radio background,
  {\it Nature Astronomy\/}, {\bf 6}, 607--617.

\bibitem[Smith et~al.(2010)Smith, Zuber, Jackson, Cavanaugh, Neumann, Riris,
  Sun, Zellar, Coltharp, Connelly, Katz, Kleyner, Liiva, Matuszeski, Mazarico,
  McGarry, Novo-Gradac, Ott, Peters, Ramos-Izquierdo, Ramsey, Rowlands,
  Schmidt, Scott, Shaw, Smith, Swinski, Torrence, Unger, Yu, \&
  Zagwodzki]{Smith2010}
Smith, D., Zuber, M., Jackson, G., Cavanaugh, J., Neumann, G., Riris, H., Sun,
  X., Zellar, R., Coltharp, C., Connelly, J., Katz, R., Kleyner, I., Liiva, P.,
  Matuszeski, A., Mazarico, E., McGarry, J., Novo-Gradac, A., Ott, M., Peters,
  C., Ramos-Izquierdo, L., Ramsey, L., Rowlands, D., Schmidt, S., Scott, V.,
  Shaw, G., Smith, J., Swinski, J., Torrence, M., Unger, G., Yu, A., \&
  Zagwodzki, T., 2010.
\newblock The lunar orbiter laser altimeter investigation on the lunar
  reconnaissance orbiter mission, {\it Space Science Reviews\/}, {\bf
  150}(1-4), 209--241.

\bibitem[Smith et~al.(1997)Smith, Zuber, Neumann, \& Lemoine]{Smith1997}
Smith, D.~E., Zuber, M.~T., Neumann, G.~A., \& Lemoine, F.~G., 1997.
\newblock Topography of the moon from the clementine lidar, {\it Journal of
  Geophysical Research: Planets\/}, {\bf 102}(E1), 1591--1611.

\bibitem[Smith et~al.(2017)Smith, Zuber, Neumann, Mazarico, LeMoine, Head,
  Lucey, Aharonson, Robinson, Sun, Torrence, Barker, Oberst, Duxbury, Mao,
  Barnouin, Jha, Rowlands, Goossens, Baker, Bauer, Gl{\"a}ser, Lemelin,
  Rosenburg, Sori, Whitten, \& McClanahan]{Smith2017}
Smith, D.~E., Zuber, M.~T., Neumann, G.~A., Mazarico, E.~M., LeMoine, F., Head,
  J.~W., Lucey, P.~G., Aharonson, O., Robinson, M.~S., Sun, X., Torrence,
  M.~H., Barker, M.~K., Oberst, J., Duxbury, T.~C., Mao, D., Barnouin, O.~S.,
  Jha, K., Rowlands, D.~D., Goossens, S., Baker, D. M.~H., Bauer, S.,
  Gl{\"a}ser, P., Lemelin, M., Rosenburg, M.~A., Sori, M.~M., Whitten, J.~L.,
  \& McClanahan, T.~P., 2017.
\newblock Summary of the results from the lunar orbiter laser altimeter after
  seven years in lunar orbit, {\it Icarus\/}, {\bf 283}, 70--91.

\bibitem[{Stappers, B. W.} et~al.(2011){Stappers, B. W.}, {Hessels, J. W. T.},
  {Alexov, A.}, {Anderson, K.}, {Coenen, T.}, {Hassall, T.}, {Karastergiou,
  A.}, {Kondratiev, V. I.}, {Kramer, M.}, {van Leeuwen, J.}, {Mol, J. D.},
  {Noutsos, A.}, {Romein, J. W.}, {Weltevrede, P.}, {Fender, R.}, {Wijers, R.
  A. M. J.}, {B\"ahren, L.}, {Bell, M. E.}, {Broderick, J.}, {Daw, E. J.},
  {Dhillon, V. S.}, {Eisl\"offel, J.}, {Falcke, H.}, {Griessmeier, J.}, {Law,
  C.}, {Markoff, S.}, {Miller-Jones, J. C. A.}, {Scheers, B.}, {Spreeuw, H.},
  {Swinbank, J.}, {ter Veen, S.}, {Wise, M. W.}, {Wucknitz, O.}, {Zarka, P.},
  {Anderson, J.}, {Asgekar, A.}, {Avruch, I. M.}, {Beck, R.}, {Bennema, P.},
  {Bentum, M. J.}, {Best, P.}, {Bregman, J.}, {Brentjens, M.}, {van de Brink,
  R. H.}, {Broekema, P. C.}, {Brouw, W. N.}, {Br\"uggen, M.}, {de Bruyn, A.
  G.}, {Butcher, H. R.}, {Ciardi, B.}, {Conway, J.}, {Dettmar, R.-J.}, {van
  Duin, A.}, {van Enst, J.}, {Garrett, M.}, {Gerbers, M.}, {Grit, T.}, {Gunst,
  A.}, {van Haarlem, M. P.}, {Hamaker, J. P.}, {Heald, G.}, {Hoeft, M.},
  {Holties, H.}, {Horneffer, A.}, {Koopmans, L. V. E.}, {Kuper, G.}, {Loose,
  M.}, {Maat, P.}, {McKay-Bukowski, D.}, {McKean, J. P.}, {Miley, G.},
  {Morganti, R.}, {Nijboer, R.}, {Noordam, J. E.}, {Norden, M.}, {Olofsson,
  H.}, {Pandey-Pommier, M.}, {Polatidis, A.}, {Reich, W.}, {R\"ottgering, H.},
  {Schoenmakers, A.}, {Sluman, J.}, {Smirnov, O.}, {Steinmetz, M.}, {Sterks, C.
  G. M.}, {Tagger, M.}, {Tang, Y.}, {Vermeulen, R.}, {Vermaas, N.}, {Vogt, C.},
  {de Vos, M.}, {Wijnholds, S. J.}, {Yatawatta, S.}, \& {Zensus,
  A.}]{Stappers2011}
{Stappers, B. W.}, {Hessels, J. W. T.}, {Alexov, A.}, {Anderson, K.}, {Coenen,
  T.}, {Hassall, T.}, {Karastergiou, A.}, {Kondratiev, V. I.}, {Kramer, M.},
  {van Leeuwen, J.}, {Mol, J. D.}, {Noutsos, A.}, {Romein, J. W.}, {Weltevrede,
  P.}, {Fender, R.}, {Wijers, R. A. M. J.}, {B\"ahren, L.}, {Bell, M. E.},
  {Broderick, J.}, {Daw, E. J.}, {Dhillon, V. S.}, {Eisl\"offel, J.}, {Falcke,
  H.}, {Griessmeier, J.}, {Law, C.}, {Markoff, S.}, {Miller-Jones, J. C. A.},
  {Scheers, B.}, {Spreeuw, H.}, {Swinbank, J.}, {ter Veen, S.}, {Wise, M. W.},
  {Wucknitz, O.}, {Zarka, P.}, {Anderson, J.}, {Asgekar, A.}, {Avruch, I. M.},
  {Beck, R.}, {Bennema, P.}, {Bentum, M. J.}, {Best, P.}, {Bregman, J.},
  {Brentjens, M.}, {van de Brink, R. H.}, {Broekema, P. C.}, {Brouw, W. N.},
  {Br\"uggen, M.}, {de Bruyn, A. G.}, {Butcher, H. R.}, {Ciardi, B.}, {Conway,
  J.}, {Dettmar, R.-J.}, {van Duin, A.}, {van Enst, J.}, {Garrett, M.},
  {Gerbers, M.}, {Grit, T.}, {Gunst, A.}, {van Haarlem, M. P.}, {Hamaker, J.
  P.}, {Heald, G.}, {Hoeft, M.}, {Holties, H.}, {Horneffer, A.}, {Koopmans, L.
  V. E.}, {Kuper, G.}, {Loose, M.}, {Maat, P.}, {McKay-Bukowski, D.}, {McKean,
  J. P.}, {Miley, G.}, {Morganti, R.}, {Nijboer, R.}, {Noordam, J. E.},
  {Norden, M.}, {Olofsson, H.}, {Pandey-Pommier, M.}, {Polatidis, A.}, {Reich,
  W.}, {R\"ottgering, H.}, {Schoenmakers, A.}, {Sluman, J.}, {Smirnov, O.},
  {Steinmetz, M.}, {Sterks, C. G. M.}, {Tagger, M.}, {Tang, Y.}, {Vermeulen,
  R.}, {Vermaas, N.}, {Vogt, C.}, {de Vos, M.}, {Wijnholds, S. J.}, {Yatawatta,
  S.}, \& {Zensus, A.}, 2011.
\newblock Observing pulsars and fast transients with lofar, {\it A\&A\/}, {\bf
  530}, A80.

\bibitem[{van Breugel} et~al.(1984){van Breugel}, {Miley}, \&
  {Heckman}]{Van1984}
{van Breugel}, W., {Miley}, G., \& {Heckman}, T., 1984.
\newblock {Studies of kiloparsec-scale, steep-spectrum radio cores. I. VLA
  maps.}, {\it \aj\/}, {\bf 89}, 5--22.

\bibitem[{van Haarlem} et~al.(2013){van Haarlem}, {Wise, M. W.}, {Gunst, A.
  W.}, {Heald, G.}, {McKean, J. P.}, {Hessels, J. W. T.}, {de Bruyn, A. G.},
  {Nijboer, R.}, {Swinbank, J.}, {Fallows, R.}, {Brentjens, M.}, {Nelles, A.},
  {Beck, R.}, {Falcke, H.}, {Fender, R.}, {H\"orandel, J.}, {Koopmans, L. V.
  E.}, {Mann, G.}, {Miley, G.}, {R\"ottgering, H.}, {Stappers, B. W.}, {Wijers,
  R. A. M. J.}, {Zaroubi, S.}, {van den Akker, M.}, {Alexov, A.}, {Anderson,
  J.}, {Anderson, K.}, {van Ardenne, A.}, {Arts, M.}, {Asgekar, A.}, {Avruch,
  I. M.}, {Batejat, F.}, {B\"ahren, L.}, {Bell, M. E.}, {Bell, M. R.}, {van
  Bemmel, I.}, {Bennema, P.}, {Bentum, M. J.}, {Bernardi, G.}, {Best, P.},
  {B\^{\i}rzan, L.}, {Bonafede, A.}, {Boonstra, A.-J.}, {Braun, R.}, {Bregman,
  J.}, {Breitling, F.}, {van de Brink, R. H.}, {Broderick, J.}, {Broekema, P.
  C.}, {Brouw, W. N.}, {Br\"uggen, M.}, {Butcher, H. R.}, {van Cappellen, W.},
  {Ciardi, B.}, {Coenen, T.}, {Conway, J.}, {Coolen, A.}, {Corstanje, A.},
  {Damstra, S.}, {Davies, O.}, {Deller, A. T.}, {Dettmar, R.-J.}, {van Diepen,
  G.}, {Dijkstra, K.}, {Donker, P.}, {Doorduin, A.}, {Dromer, J.}, {Drost, M.},
  {van Duin, A.}, {Eisl\"offel, J.}, {van Enst, J.}, {Ferrari, C.}, {Frieswijk,
  W.}, {Gankema, H.}, {Garrett, M. A.}, {de Gasperin, F.}, {Gerbers, M.}, {de
  Geus, E.}, {Grie\ss{}meier, J.-M.}, {Grit, T.}, {Gruppen, P.}, {Hamaker, J.
  P.}, {Hassall, T.}, {Hoeft, M.}, {Holties, H. A.}, {Horneffer, A.}, {van der
  Horst, A.}, {van Houwelingen, A.}, {Huijgen, A.}, {Iacobelli, M.}, {Intema,
  H.}, {Jackson, N.}, {Jelic, V.}, {de Jong, A.}, {Juette, E.}, {Kant, D.},
  {Karastergiou, A.}, {Koers, A.}, {Kollen, H.}, {Kondratiev, V. I.},
  {Kooistra, E.}, {Koopman, Y.}, {Koster, A.}, {Kuniyoshi, M.}, {Kramer, M.},
  {Kuper, G.}, {Lambropoulos, P.}, {Law, C.}, {van Leeuwen, J.}, {Lemaitre,
  J.}, {Loose, M.}, {Maat, P.}, {Macario, G.}, {Markoff, S.}, {Masters, J.},
  {McFadden, R. A.}, {McKay-Bukowski, D.}, {Meijering, H.}, {Meulman, H.},
  {Mevius, M.}, {Middelberg, E.}, {Millenaar, R.}, {Miller-Jones, J. C. A.},
  {Mohan, R. N.}, {Mol, J. D.}, {Morawietz, J.}, {Morganti, R.}, {Mulcahy, D.
  D.}, {Mulder, E.}, {Munk, H.}, {Nieuwenhuis, L.}, {van Nieuwpoort, R.},
  {Noordam, J. E.}, {Norden, M.}, {Noutsos, A.}, {Offringa, A. R.}, {Olofsson,
  H.}, {Omar, A.}, {Orr\'u, E.}, {Overeem, R.}, {Paas, H.}, {Pandey-Pommier,
  M.}, {Pandey, V. N.}, {Pizzo, R.}, {Polatidis, A.}, {Rafferty, D.},
  {Rawlings, S.}, {Reich, W.}, {de Reijer, J.-P.}, {Reitsma, J.}, {Renting, G.
  A.}, {Riemers, P.}, {Rol, E.}, {Romein, J. W.}, {Roosjen, J.}, {Ruiter, M.},
  {Scaife, A.}, {van der Schaaf, K.}, {Scheers, B.}, {Schellart, P.},
  {Schoenmakers, A.}, {Schoonderbeek, G.}, {Serylak, M.}, {Shulevski, A.},
  {Sluman, J.}, {Smirnov, O.}, {Sobey, C.}, {Spreeuw, H.}, {Steinmetz, M.},
  {Sterks, C. G. M.}, {Stiepel, H.-J.}, {Stuurwold, K.}, {Tagger, M.}, {Tang,
  Y.}, {Tasse, C.}, {Thomas, I.}, {Thoudam, S.}, {Toribio, M. C.}, {van der
  Tol, B.}, {Usov, O.}, {van Veelen, M.}, {van der Veen, A.-J.}, {ter Veen,
  S.}, {Verbiest, J. P. W.}, {Vermeulen, R.}, {Vermaas, N.}, {Vocks, C.},
  {Vogt, C.}, {de Vos, M.}, {van der Wal, E.}, {van Weeren, R.}, {Weggemans,
  H.}, {Weltevrede, P.}, {White, S.}, {Wijnholds, S. J.}, {Wilhelmsson, T.},
  {Wucknitz, O.}, {Yatawatta, S.}, {Zarka, P.}, {Zensus, A.}, \& {van Zwieten,
  J.}]{Haarlem2013}
{van Haarlem}, {Wise, M. W.}, {Gunst, A. W.}, {Heald, G.}, {McKean, J. P.},
  {Hessels, J. W. T.}, {de Bruyn, A. G.}, {Nijboer, R.}, {Swinbank, J.},
  {Fallows, R.}, {Brentjens, M.}, {Nelles, A.}, {Beck, R.}, {Falcke, H.},
  {Fender, R.}, {H\"orandel, J.}, {Koopmans, L. V. E.}, {Mann, G.}, {Miley,
  G.}, {R\"ottgering, H.}, {Stappers, B. W.}, {Wijers, R. A. M. J.}, {Zaroubi,
  S.}, {van den Akker, M.}, {Alexov, A.}, {Anderson, J.}, {Anderson, K.}, {van
  Ardenne, A.}, {Arts, M.}, {Asgekar, A.}, {Avruch, I. M.}, {Batejat, F.},
  {B\"ahren, L.}, {Bell, M. E.}, {Bell, M. R.}, {van Bemmel, I.}, {Bennema,
  P.}, {Bentum, M. J.}, {Bernardi, G.}, {Best, P.}, {B\^{\i}rzan, L.},
  {Bonafede, A.}, {Boonstra, A.-J.}, {Braun, R.}, {Bregman, J.}, {Breitling,
  F.}, {van de Brink, R. H.}, {Broderick, J.}, {Broekema, P. C.}, {Brouw, W.
  N.}, {Br\"uggen, M.}, {Butcher, H. R.}, {van Cappellen, W.}, {Ciardi, B.},
  {Coenen, T.}, {Conway, J.}, {Coolen, A.}, {Corstanje, A.}, {Damstra, S.},
  {Davies, O.}, {Deller, A. T.}, {Dettmar, R.-J.}, {van Diepen, G.}, {Dijkstra,
  K.}, {Donker, P.}, {Doorduin, A.}, {Dromer, J.}, {Drost, M.}, {van Duin, A.},
  {Eisl\"offel, J.}, {van Enst, J.}, {Ferrari, C.}, {Frieswijk, W.}, {Gankema,
  H.}, {Garrett, M. A.}, {de Gasperin, F.}, {Gerbers, M.}, {de Geus, E.},
  {Grie\ss{}meier, J.-M.}, {Grit, T.}, {Gruppen, P.}, {Hamaker, J. P.},
  {Hassall, T.}, {Hoeft, M.}, {Holties, H. A.}, {Horneffer, A.}, {van der
  Horst, A.}, {van Houwelingen, A.}, {Huijgen, A.}, {Iacobelli, M.}, {Intema,
  H.}, {Jackson, N.}, {Jelic, V.}, {de Jong, A.}, {Juette, E.}, {Kant, D.},
  {Karastergiou, A.}, {Koers, A.}, {Kollen, H.}, {Kondratiev, V. I.},
  {Kooistra, E.}, {Koopman, Y.}, {Koster, A.}, {Kuniyoshi, M.}, {Kramer, M.},
  {Kuper, G.}, {Lambropoulos, P.}, {Law, C.}, {van Leeuwen, J.}, {Lemaitre,
  J.}, {Loose, M.}, {Maat, P.}, {Macario, G.}, {Markoff, S.}, {Masters, J.},
  {McFadden, R. A.}, {McKay-Bukowski, D.}, {Meijering, H.}, {Meulman, H.},
  {Mevius, M.}, {Middelberg, E.}, {Millenaar, R.}, {Miller-Jones, J. C. A.},
  {Mohan, R. N.}, {Mol, J. D.}, {Morawietz, J.}, {Morganti, R.}, {Mulcahy, D.
  D.}, {Mulder, E.}, {Munk, H.}, {Nieuwenhuis, L.}, {van Nieuwpoort, R.},
  {Noordam, J. E.}, {Norden, M.}, {Noutsos, A.}, {Offringa, A. R.}, {Olofsson,
  H.}, {Omar, A.}, {Orr\'u, E.}, {Overeem, R.}, {Paas, H.}, {Pandey-Pommier,
  M.}, {Pandey, V. N.}, {Pizzo, R.}, {Polatidis, A.}, {Rafferty, D.},
  {Rawlings, S.}, {Reich, W.}, {de Reijer, J.-P.}, {Reitsma, J.}, {Renting, G.
  A.}, {Riemers, P.}, {Rol, E.}, {Romein, J. W.}, {Roosjen, J.}, {Ruiter, M.},
  {Scaife, A.}, {van der Schaaf, K.}, {Scheers, B.}, {Schellart, P.},
  {Schoenmakers, A.}, {Schoonderbeek, G.}, {Serylak, M.}, {Shulevski, A.},
  {Sluman, J.}, {Smirnov, O.}, {Sobey, C.}, {Spreeuw, H.}, {Steinmetz, M.},
  {Sterks, C. G. M.}, {Stiepel, H.-J.}, {Stuurwold, K.}, {Tagger, M.}, {Tang,
  Y.}, {Tasse, C.}, {Thomas, I.}, {Thoudam, S.}, {Toribio, M. C.}, {van der
  Tol, B.}, {Usov, O.}, {van Veelen, M.}, {van der Veen, A.-J.}, {ter Veen,
  S.}, {Verbiest, J. P. W.}, {Vermeulen, R.}, {Vermaas, N.}, {Vocks, C.},
  {Vogt, C.}, {de Vos, M.}, {van der Wal, E.}, {van Weeren, R.}, {Weggemans,
  H.}, {Weltevrede, P.}, {White, S.}, {Wijnholds, S. J.}, {Wilhelmsson, T.},
  {Wucknitz, O.}, {Yatawatta, S.}, {Zarka, P.}, {Zensus, A.}, \& {van Zwieten,
  J.}, 2013.
\newblock Lofar: The low-frequency array, {\it A\&A\/}, {\bf 556}, A2.

\bibitem[Wessel et~al.(2019)Wessel, Luis, Uieda, Scharroo, Wobbe, Smith, \&
  Tian]{Wessel2019}
Wessel, P., Luis, J.~F., Uieda, L., Scharroo, R., Wobbe, F., Smith, W. H.~F.,
  \& Tian, D., 2019.
\newblock The generic mapping tools version 6, {\it Geochemistry, Geophysics,
  Geosystems\/}, {\bf 20}(11), 5556--5564.

\bibitem[Wijkman(1982)]{Wijkman1982}
Wijkman, P.~M., 1982.
\newblock Managing the global commons, {\it International Organization\/}, {\bf
  36}(3), 511–536.

\bibitem[Williams(2021)]{Williams2021}
Williams, D.~R., 2021.
\newblock Moon fact sheet.

\end{thebibliography}

\appendix
\section{Topographic Maps}
\label{app:maps}
Here the measured elevation and pulse width and the derived slope and roughness maps are presented for all eight candidate sites on the lunar far-side.

\begin{figure*}
    \centering
    \includegraphics[height=0.9\textheight]{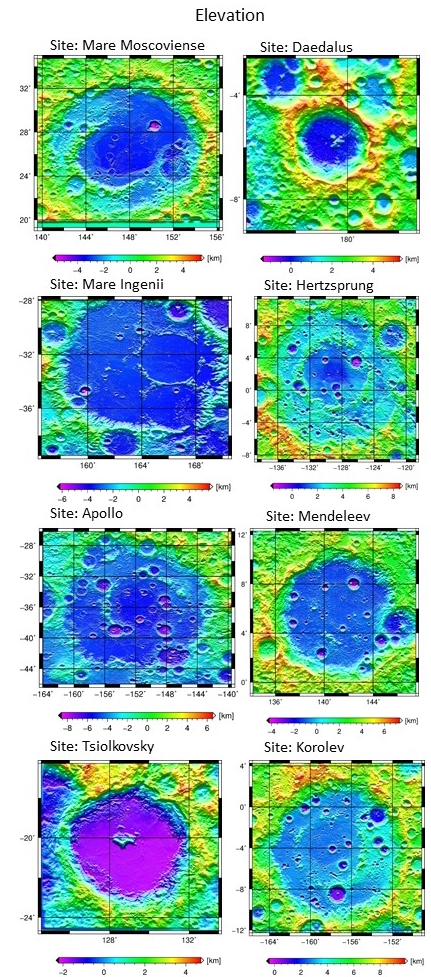}
    \caption{Elevation maps of all eight sites from LOLA data. Elevation in km is shown by a colour scale. The site names are labelled at the top left of each map. A 4°$x$4° ($122 x 122$ km) grid is overlaid on each site map, showing their relative sizes.}
    \label{fig:Elevation}
\end{figure*}
\begin{figure*}
    \centering
    \includegraphics[height=0.9\textheight]{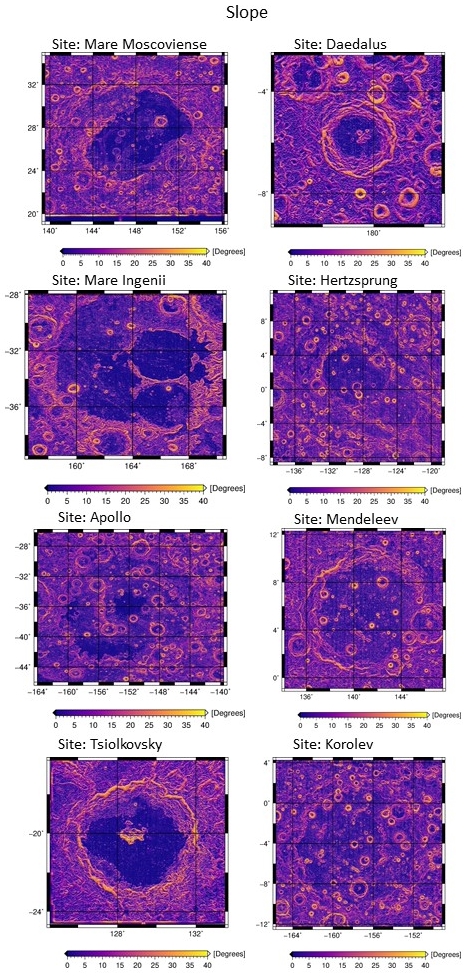}
    \caption{Slope maps of all eight sites from LOLA data. Slope in degrees is shown by a colour scale. The site names are labelled at the top left of each map. A 4°$x$4° ($122 x 122$ km) grid is overlaid on each site map, showing their relative sizes.}
    \label{fig:Slope}
\end{figure*}
\begin{figure*}
    \centering
    \includegraphics[height=0.9\textheight]{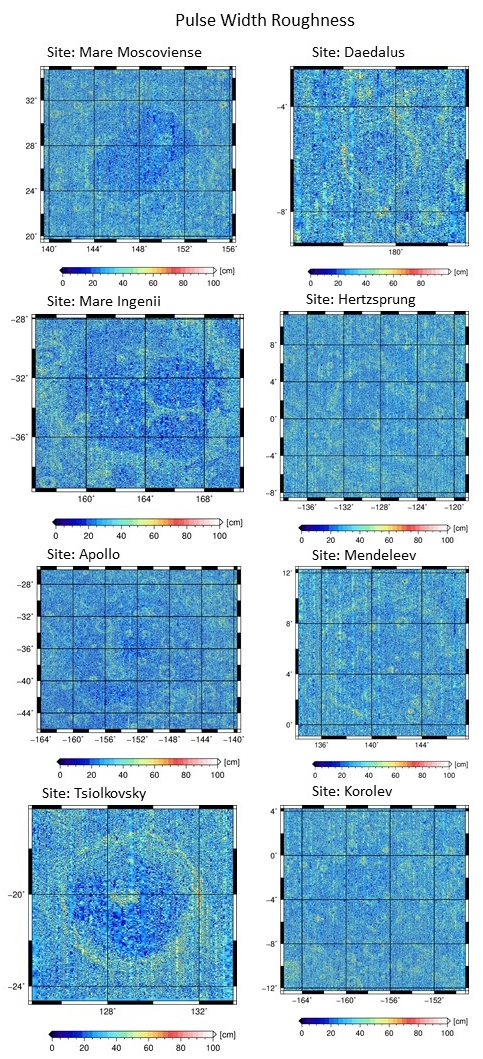}
    \caption{Pulse width roughness maps of all eight sites measured by LOLA. Roughness in cm is shown by a colour scale. The site names are labelled at the top left of each map. A 4°$x$4° ($122 x 122$ km) grid is overlaid on each site map, showing their relative sizes. The pulse width measurements are not used in this study because of the lack of LOLA data due to a thermal anomaly.}
    \label{fig:PW}
\end{figure*}
\begin{figure*}
    \centering
    \includegraphics[height=0.9\textheight]{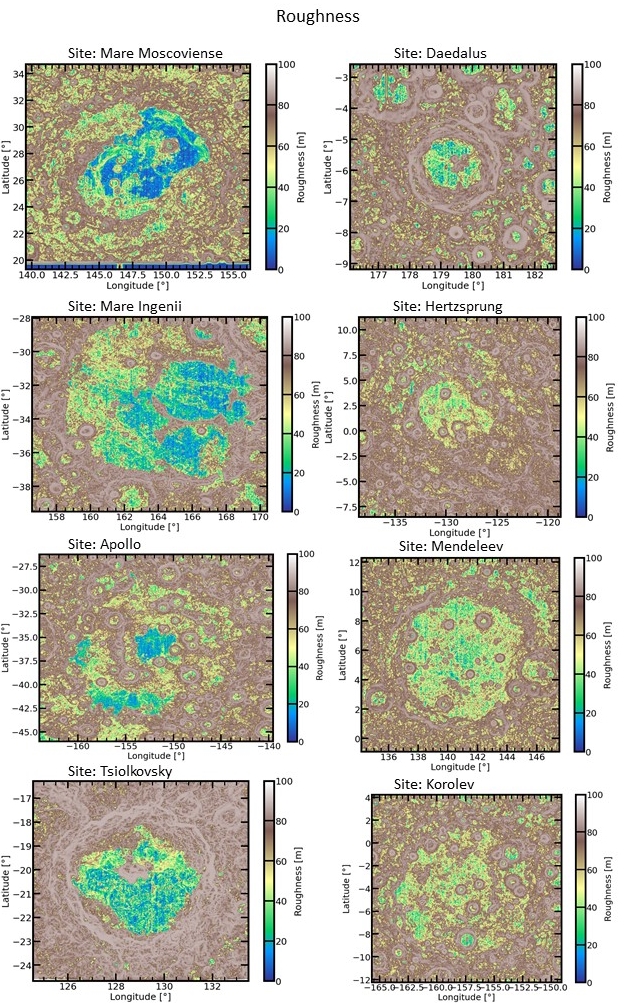}
    \caption{RMS Roughness maps of all eight sites derived from LOLA data. Roughness in metres is shown by a colour scale. The site names are labelled at the top left of each map. Blue areas can be traversed by a rover and brown areas would not be traversable by a rover.}
    \label{fig:RMS}
\end{figure*}
\begin{figure*}
    \centering
    \includegraphics[height=0.9\textheight]{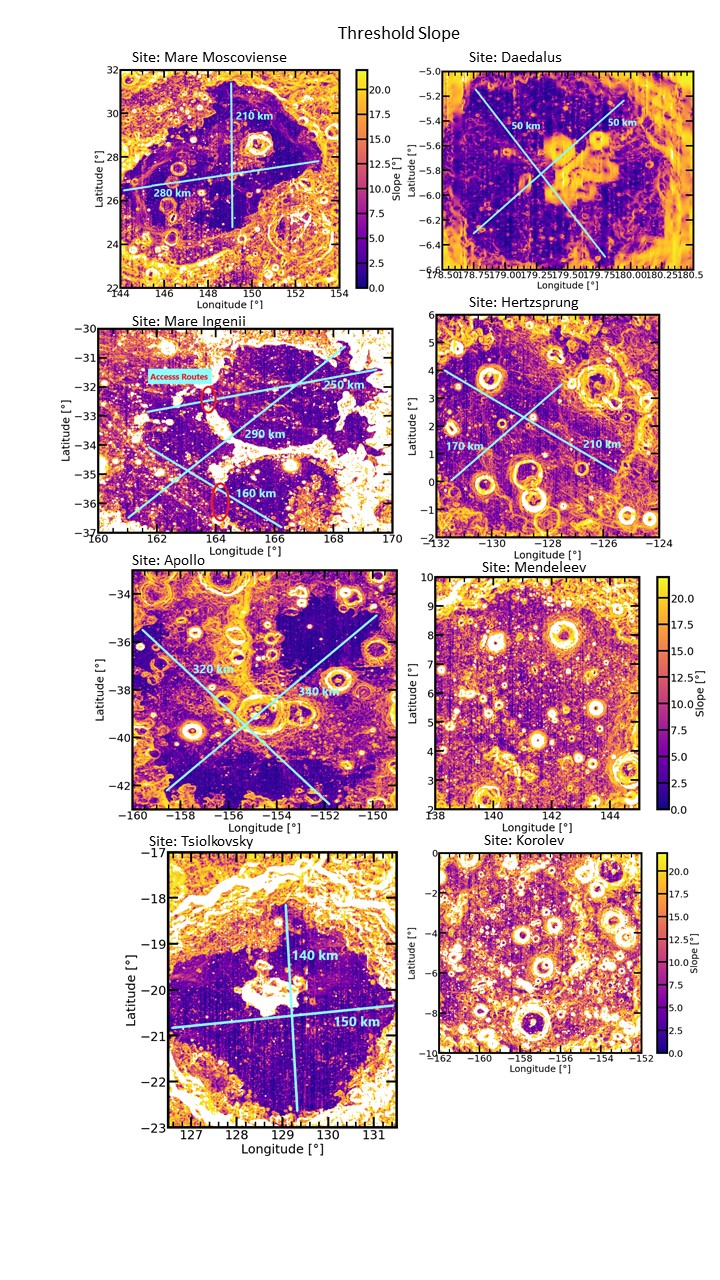}
    \caption{22° slope threshold maps of all lunar sites of interest derived from LOLA data. The slope in degrees is shown by a colour scale. The site names are labelled at the top left of each map. Terrain not traversable by a rover is shown by white space, where slopes are greater than 22°. Potential baselines are shown in cyan and their lengths are labelled in km, note that the rover path may be significantly longer than the baseline illustrated.}
    \label{fig:Threshold}
\end{figure*}

\bsp	
\label{lastpage}
\end{document}